\documentclass{aa}  

\bibpunct{(}{)}{;}{a}{}{,}

\usepackage{graphicx, graphics}
\usepackage{txfonts}
\usepackage{longtable}

\begin{document} 

   \title{Physical and chemical fingerprint of protostellar disc formation}


   \author{E. Artur de la Villarmois
          \inst{1} \and
          J. K. J{\o}rgensen
          \inst{1} \and
          L. E. Kristensen
          \inst{1} \and
         E. A. Bergin
          \inst{2} \and
          D. Harsono
          \inst{3} \and
           N. Sakai
          \inst{4} \and
           E. F. van Dishoeck
          \inst{3, 5} \and
          S. Yamamoto
          \inst{6}
          }

\institute{Niels Bohr Institute $\&$ Centre for Star and Planet Formation, University of Copenhagen, 
   {\O}ster Voldgade 5--7, 1350 Copenhagen K., Denmark\\
              \email{elizabeth.artur@nbi.ku.dk}
\and Department of Astronomy, University of Michigan, 311 West Hall, 1085 S. University Ave, Ann Arbor, MI 48109, USA
\and Leiden Observatory, Leiden University, PO Box 9513, NL-2300 RA Leiden, the Netherlands
\and The Institute of Physical and Chemical Research (RIKEN), 2-1 Hirosawa, Wako-shi, Saitama 351-0198, Japan
\and Max-Planck-Institut f$\ddot{\mathrm{u}}$r extraterrestrische Physik, Giessenbachstra{\ss}e 1, 85748, Garching bei M$\ddot{\mathrm{u}}$nchen, Germany
\and Department of Physics, The University of Tokyo, Bunkyo-ku, Tokyo 113-0033, Japan
             }

  \abstract
    {The structure and composition of emerging planetary systems are likely strongly influenced by their natal environment within the protoplanetary disc at the time when the star is still gaining mass. It is therefore essential to identify and study the physical processes at play in the gas and dust close to young protostars and investigate the chemical composition of the material that is inherited from the parental cloud.}
    {The purpose of this paper is to explore and compare the physical and chemical structure of Class I low-mass protostellar sources on protoplanetary disc scales.}
    {We present a study of the dust and gas emission towards a representative sample of 12 Class I protostars from the Ophiuchus molecular cloud with the Atacama Large Millimeter/submillimeter Array (ALMA). The continuum at 0.87~mm and molecular transitions from C$^{17}$O, C$^{34}$S, H$^{13}$CO$^{+}$, CH$_{3}$OH, SO$_{2}$ , and C$_{2}$H were observed at high angular resolution (0$\farcs$4, $\sim$60~au diameter) towards each source. The spectrally and spatially resolved maps reveal the kinematics and the spatial distribution of each species. Moreover, disc and stellar masses are estimated from the continuum flux and position-velocity diagrams, respectively.}
    {Six of the sources show disc-like structures in C$^{17}$O, C$^{34}$S , or H$^{13}$CO$^{+}$ emission. Towards the more luminous sources, compact emission and large line widths are seen for transitions of SO$_2$ that probe warm gas (\textit{E$_\mathrm{u}$}~$\sim$200~K). In contrast, C$^{17}$O emission is detected towards the least evolved and less luminous systems. No emission of CH$_3$OH is detected towards any of the continuum peaks, indicating an absence of warm CH$_3$OH gas towards these sources.}    
     {A trend of increasing stellar mass is observed as the envelope mass decreases. In addition, a power-law relation is seen between the stellar mass and the bolometric luminosity, corresponding to a mass accretion rate of (2.4~$\pm$~0.6)~$\times$~10$^{-7}$ M$_{\odot}$~year$^{-1}$ for the Class I sources, with a minimum and maximum value of 7.5~$\times$~10$^{-8}$ and 7.6~$\times$~10$^{-7}$ M$_{\odot}$~year$^{-1}$, respectively. This mass accretion rate is lower than the expected value if the accretion is constant in time and rather points to a scenario of accretion occurring in bursts. The differentiation between C$^{17}$O and SO$_{2}$ suggests that they trace different physical components: C$^{17}$O traces the densest and colder regions of the disc-envelope system, while SO$_{2}$ may be associated with regions of higher temperature, such as accretion shocks. The lack of warm CH$_{3}$OH emission suggests that there is no hot-core-like region around any of the sources and that the CH$_{3}$OH column density averaged over the disc is low. Finally, the combination of bolometric temperature and luminosity may indicate an evolutionary trend of chemical composition during these early stages.}


   \maketitle

\section{Introduction}

The formation and evolution of protoplanetary discs are fundamental in the process of low-mass star formation and to understand how our own solar system formed. The chemical complexity of the protoplanetary disc is established either by the material that is inherited from the envelope or from processed material within the disc, or a combination of the two \citep[e.g.][]{Pontoppidan2014, Drozdovskaya2018}, providing the initial chemical conditions for planet formation. However, the physical and chemical processes at play on small scales ($\leq$ 500~au) in low-mass protostars are still not well understood.

The dynamical evolution of the system affects the mass distribution, where material from the inner envelope falls towards the disc, the disc accretes material onto the central protostar, and part of this material is ejected through outflows \citep{Terebey1984, Shu1993, Hartmann1998}. Class~I sources are associated with the formation and evolution of circumstellar discs \citep[e.g.][]{Sheehan2017, Yoo2017} and still have a significant contribution from the envelope material \citep{Robitaille2006}. Therefore, Class~I sources act as a bridge between the deeply embedded Class~0 sources and the protoplanetary discs that are associated with Class~II sources. The evolutionary sequence, Class~0 - Class~I - Class~II, still has puzzling questions such as when and how quickly the envelope dissipates, how early discs form and how quickly they grow in mass and size, and how material accretes from the disc onto the protostar. The presence of knots or bullets in outflows \citep[e.g.][]{Reipurth1989, Arce2013} and the low luminosity of protostars compared to models \citep{Kenyon1995, Evans2009, Dunham2012} suggests that the mass accretion rates may vary with time. Direct measurement of the mass accretion rate is extremely challenging for embedded protostars, but an approximate estimate can be obtained from the accretion luminosity equation \citep[e.g.][]{Kenyon1995, White2007, Dunham2014a, Mottram2017}. The changes in accretion, and thus in luminosity, may have significant consequences in the chemistry and further evolution of the system \citep[e.g.][]{Jorgensen2013, Jorgensen2015, Frimann2016a}.    

The complex environment in which protoplanetary discs form and evolve is exposed to large variations in temperature (tens to hundreds of K) and density (10$^{5}$$-$10$^{13}$~cm$^{-3}$), which leave strong chemical signatures and make molecules excellent diagnostics of the physical conditions and processes. Observed different molecular species that trace different physics, such as disc tracers \citep[such as $^{13}$CO and C$^{18}$O; e.g.][]{Harsono2014}, warm gas tracers \citep[such as CH$_{3}$OH; e.g.][]{Jorgensen2013}, and shock tracers \citep[such as SO; e.g.][]{Sakai2014}, are essential in order to understand the different physical and chemical processes involved at disc scales. Previous studies of molecular line emission towards Class~I sources were focused on characterising the gas kinematics, or the chemistry of single sources or of a few sources \citep[e.g.][]{Jorgensen2009, Harsono2014}, which makes it difficult to compare them. In addition, little is known about the evolution of the molecular content as a function of physics in these stages.

With the high sensitivity and angular resolution of the Atacama Large Millimeter/submillimeter Array (ALMA), it is becoming possible to resolve disc scales (10$-$50~au towards nearby star-forming regions) and study the physics and chemistry of these environments. Disc rotation, when detectable, is a strong tool for determining protostellar masses \citep[e.g.][]{Jorgensen2009, Harsono2014, Yen2014, Yen2015}. A comparison between the chemistry and the physical parameters of protostars at disc scales and in a more statistical way is therefore now becoming possible. 

One of the closest low-mass star-forming regions is the Ophiuchus molecular cloud \citep{Wilking2008}, with a distance (\textit{d}) of 139~$\pm$~6~pc \citep{Mamajek2008}. It is associated with protostellar sources at different evolutionary stages, which makes this star-forming region an excellent laboratory for the study of low-mass star formation and the discs of low-mass stars.  

We present ALMA observations of a representative sample of 12 Class~I sources in the Ophiuchus molecular cloud. The observations include continuum emission at 0.87~mm and molecular lines that trace different components of the star-forming environment: C$^{17}$O, C$^{34}$S, H$^{13}$CO$^{+}$, CH$_{3}$OH, SO$_{2}$, and C$_{2}$H. Section 2 describes the observational procedure, the data calibration, the source properties, and the molecular transitions that are covered. The results are presented in Sect. 3, where moment 0 and 1 maps are shown for each molecular transition. Section 4 presents an analysis of the mass evolution, with a comparison between disc masses, stellar masses, envelope masses, and bolometric luminosities, following an estimate of the mass accretion rate. Section 5 describes the chemical evolution, focusing on the lack of warm CH$_{3}$OH detection, the compact and broad SO$_{2}$ emission, and the chemical trend observed within the sources. Finally, Sect. 6 summarises our main findings.

\section{Observations}

In order to address the physical and chemical properties of discs in their earliest stages, a sample of 12 Class~I protostars in Ophiuchus was observed. The sources are all well characterised through large-scale mid-infrared (\textit{Spitzer}) and submillimetre (SCUBA) surveys \citep{Jorgensen2008}, and were selected in order to include protostars with bolometric temperatures (\textit{T$_\mathrm{bol}$}) below 400~K. In addition, the sample covers a wide range of bolometric luminosities (\textit{L$_\mathrm{bol}$}), from 0.03 to 18~L$_{\odot}$, and envelope masses from about 0.05~M$_{\odot}$ up to 0.3~M$_{\odot}$ (Table~\ref{table:parameters}). With this approach, the sample is a representative set for exploring the physical and chemical structures of Class~I sources.

The 12 sources were observed with ALMA on four occasions, between 2015 May 21 and June 5 (program code: 2013.1.00955.S; PI: Jes J{\o}rgensen). At the time of the observations, 36 antennas were available in the array (37 for the June 5 observations), providing baselines between 21 and 556 metres (784 metres for the June 5 observations) and a maximum angular scale of $\sim$18$\arcsec$. Each of the four sessions provided an on-source time of 43 minutes in total for the 12 different sources (i.e. each source was observed for approximately 15~minutes in total). The observed sources are listed in Table~\ref{table:parameters}, with their physical properties and other common identifiers.

\begin{figure*}[t]
   \centering
      \includegraphics[width=.7\textwidth]{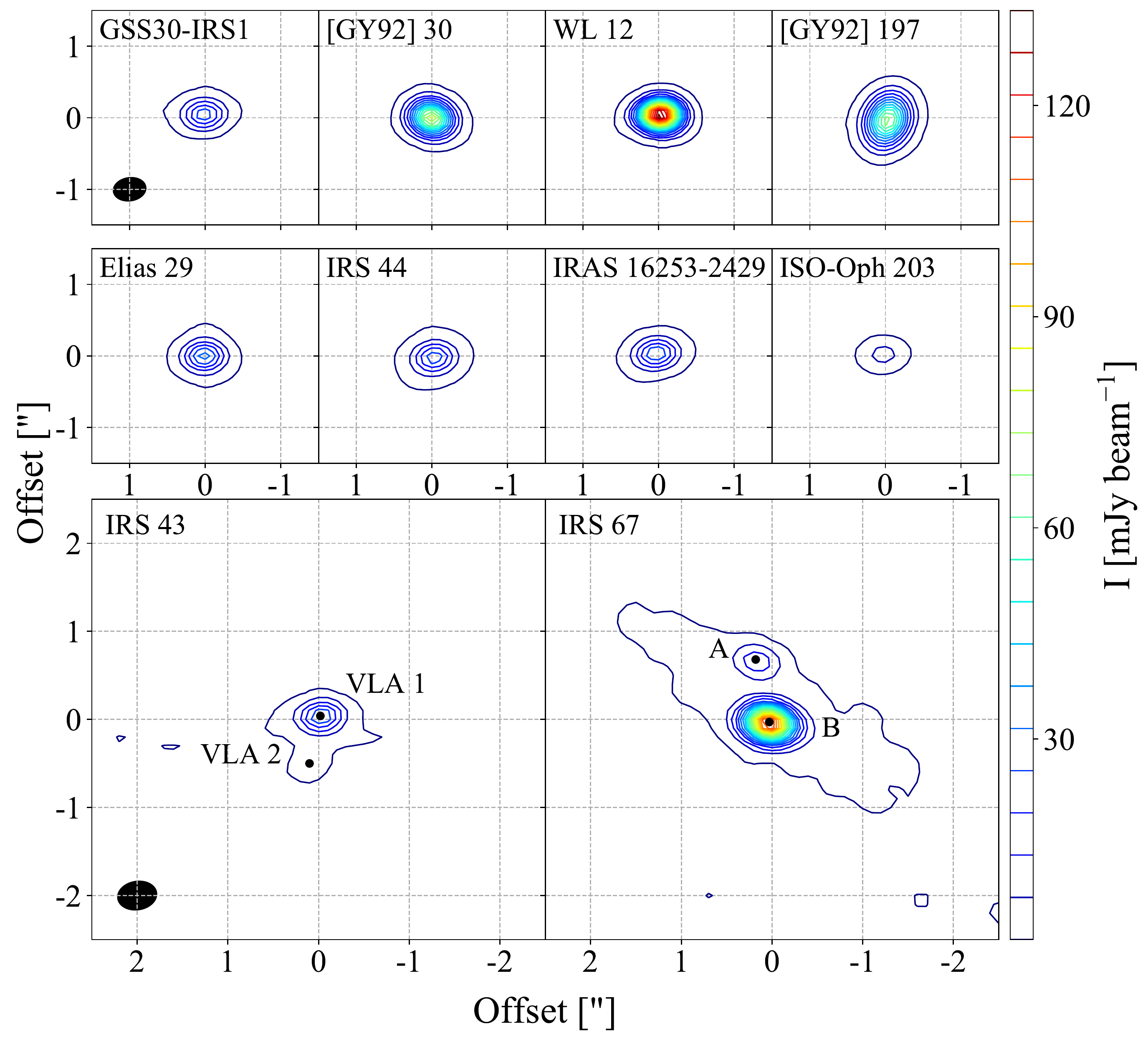}
      \caption[]{\label{fig:Continuum}
      Continuum emission of the detected sources above 5$\sigma$ ($\sigma$~=~0.3 mJy~beam$^{-1}$). The contours start at 5$\sigma$ and follow a step of 20$\sigma$. The typical synthesised beam is represented by the black filled ellipse in the upper and lower left panels. The (0,0) position represents the position fitted with a 2D Gaussian (see Table~\ref{table:coordinates}). The lower panels show the binary systems with the common identifiers of both components.
   }
\end{figure*}

\begin{table*}[t]
\caption{Properties of the observed sources: infrared position, bolometric temperature (\textit{T$_\mathrm{bol}$}), bolometric luminosity (\textit{L$_\mathrm{bol}$}), envelope mass (\textit{M$_\mathrm{env}$}), and other common identifiers.}
\label{table:parameters}
\centering
\begin{tabular}{l c c r r l l}
        \hline\hline
                 Source & \multicolumn{2}{c}{Infrared position $^{a}$} & \textit{T$_\mathrm{bol}$} $^{b}$ & \textit{L$_\mathrm{bol}$ $^{b}$} & \textit{M$_\mathrm{env}$} $^{c}$ & Other common identifiers       \\
                             & RA [J2000.0] & Dec [J2000.0]      & [K]                                          & [L$_{\odot}$]                              &   [M$_{\odot}$]                                  &                                         \\
        \hline
                J162614.6                       & 16 26 14.63   & $-$24 25 07.5    & 7             & 0.03  & 0.095 & [EDJ2009] 800                                 \\      
                GSS30-IRS1              & 16 26 21.35   & $-$24 23 04.3 & 250     & 11.00 & 0.15  & Oph-emb 8, [GY92] 6,  Elias 21                \\\relax
                [GY92] 30                       & 16 26 25.46   & $-$24 23 01.3    & 200   & 0.12  & 0.27  & Oph-emb 9                                             \\
                WL 12                   & 16 26 44.19   & $-$24 34 48.4 & 380     & 1.40  & 0.076 & VSSG 30, [GY92] 111                           \\
                IRAS 16238-2428 & 16 26 59.10   & $-$24 35 03.3 & 120   & 1.70    & 0.045 & [EDJ2009] 856                                 \\\relax
                [GY92] 197              & 16 27 05.24   & $-$24 36 29.6 & 120     & 0.18  & 0.16  & Oph-emb 6, LFAM 26                            \\ 
                Elias 29                        & 16 27 09.40   & $-$24 37 18.6    & 420   & 18.00 & 0.045 & Oph-emb 16, [GY92] 214                        \\
                IRS 43 $^{d}$           & 16 27 26.91   & $-$24 40 50.7 & 300     & 3.30  & 0.13  & Oph-emb 14, [GY92] 265, YLW 15                \\
                IRS 44                  & 16 27 27.98   & $-$24 39 33.4 & 280     & 7.10  & 0.057 & Oph-emb 13, [GY92] 269, YLW 16                \\ 
                IRAS 16253-2429 & 16 28 21.61   & $-$24 36 23.4 & 36            & 0.24    & 0.15  & Oph-emb 1, MMS 126                            \\
                ISO-Oph 203             & 16 31 52.45   & $-$24 55 36.2 & 330     & 0.15  & 0.072 & Oph-emb 15                                            \\
                IRS 67 $^{d}$           & 16 32 00.99   & $-$24 56 42.6 & 180     & 2.80  & 0.078 & Oph-emb 10, L1689S1 3                 \\      
        \hline
\end{tabular}
\tablefoot{$^{(a)}$ Position of \textit{Spitzer} sources from the c2d catalog \citep{Evans2009}.$^{(b)}$ From \cite{Dunham2015}. $^{(c)}$ Envelope mass from models of sub-millimetre and IR data following description in \cite{Jorgensen2009}. $^{(d)}$ Binary system.}
\end{table*}

The observations covered five different line settings, and the choice of species was made specifically to trace different aspects of the structure of protostars. Lines of C$^{17}$O, H$^{13}$CO$^{+}$ and C$^{34}$S were targeted as tracers of disc kinematics, while SO$_{2}$ and CH$_{3}$OH are expected to trace the warm chemistry in the inner envelope or disc, and C$_{2}$H is commonly associated with outer-envelope regions. Two spectral windows consist of 960~channels each with 122.07~kHz (0.11~km~s$^{-1}$) spectral resolution centred on C$^{17}$O \textit{J}=3$-$2 and C$^{34}$S \textit{J}=7$-$6, while the other three contain 1920~channels each with 244.14~kHz (0.22~km~s$^{-1}$) spectral resolution centred on H$^{13}$CO$^+$ \textit{J}=4$-$3, the CH$_3$OH \textit{J$_{k}$}=7$_k$$-$6$_k$ branch at 338.4~GHz, and the CH$_3$CN 14$-$13 branch at 349.1~GHz. The latter two settings also pick up SO$_2$ and C$_2$H transitions. The observed molecular transitions and their parameters are summarised in Table~\ref{table:observations}.

\begin{table*}[t]
        \caption{Spectral setup and parameters of the detected molecular transitions.}
        \label{table:observations}
        \centering
        \begin{tabular}{l c c r c}
                \hline\hline
                Molecular transition & Frequency $^{a}$ & \textit{A$_{ij}$} $^{a}$ & \textit{E$_\mathrm{u}$} $^{a}$ & \textit{n$_\mathrm{crit}$} $^{b}$\\
                  & [GHz] & [s$^{-1}$] &  [K] & [cm$^{-3}$] \\
                \hline
                C$^{17}$O \textit{J}=3$-$2                                                                      & 337.06110       & 2.3 $\times$ 10$^{-6}$        & 32            & 3.5 $\times$ 10$^{4}$        \\
                C$^{34}$S \textit{J}=7$-$6                                                                      & 337.39646       & 8.4 $\times$ 10$^{-4}$        & 50            & 1.3 $\times$ 10$^{7}$        \\
                SO$_{2}$ \textit{J}$_\mathrm{K_{A}K_{C}}$=18$_{4,14}$$-$18$_{3,15}$         & 338.30599     & 3.3 $\times$ 10$^{-4}$        & 197   & 5.9 $\times$ 10$^{7}$        \\               
                CH$_{3}$OH \textit{J}$_\mathrm{K}$=7$_{-1}$$-$6$_{-1}$ E                         & 338.34463     & 1.7 $\times$ 10$^{-4}$        & 70              & 2.0 $\times$ 10$^{6}$ \\               
                CH$_{3}$OH \textit{J}$_\mathrm{K}$=7$_{0}$$-$6$_{0}$ A$^{+}$                 & 338.40868     & 1.7 $\times$ 10$^{-4}$        & 65             & 2.8 $\times$ 10$^{7}$ \\
                H$^{13}$CO$^{+}$ \textit{J}=4$-$3                                                               & 346.99835       & 3.3 $\times$ 10$^{-3}$        & 42            & 8.5 $\times$ 10$^{6}$        \\
                C$_{2}$H \textit{N}=4$-$3, \textit{J}=9/2$-$7/2, \textit{F}=5$-$4                 & 349.33774     & 1.3 $\times$ 10$^{-4}$        & 42             & 2.2 $\times$ 10$^{7}$         \\               
                C$_{2}$H \textit{N}=4$-$3, \textit{J}=9/2$-$7/2, \textit{F}=4$-$3                 & 349.39934     & 1.3 $\times$ 10$^{-4}$        & 42             & 2.3 $\times$ 10$^{7}$         \\
                \hline
        \end{tabular}
        \tablefoot{$^{(a)}$  Values from the CDMS database \citep{Muller2001}. $^{(b)}$ Calculated values for a kinetic temperature of 30~K and collisional rates from the Leiden Atomic and Molecular Database \citep[LAMDA; ][]{Schoier2005}. The collisional rates of specific species were taken from the following sources: C$^{17}$O from \cite{Yang2010}, C$^{34}$S from \cite{Lique2006}, SO$_{2}$ from \cite{Balanca2016}, CH$_{3}$OH from \cite{Rabli2010}, H$^{13}$CO$^{+}$ from \cite{Flower1999}, and C$_{2}$H from \cite{Spielfiedel2012}.}
\end{table*}

\begin{table*}[t]
        \caption{Results of 2D Gaussian fits towards the continuum peak of our Oph sources.}
        \label{table:coordinates}
        \centering
        \begin{tabular}{l c c l l r r}
        \hline\hline
                \multicolumn{1}{c}{Source} & RA & Dec   & \multicolumn{1}{c}{Size $^{a}$} & \multicolumn{1}{c}{PA}         & \multicolumn{1}{c}{\textit{F$_\mathrm{0.87 mm}$}}  & \multicolumn{1}{c}{\textit{S$_\mathrm{0.87 mm}$} (0\farcs4) $^{b}$}         \\
                                                & [J2000.0]     & [J2000.0]     & \multicolumn{1}{c}{[$\arcsec$]} & \multicolumn{1}{c}{[$\degr$]} & \multicolumn{1}{c}{[mJy]}         & \multicolumn{1}{c}{[mJy beam$^{-1}$]} \\
        \hline
                GSS30-IRS1              & 16 26 21.358          & -24 23 04.85           & 0.19~$\pm$~0.02~$\times$~0.12~$\pm$~0.01              & \enspace96~$\pm$~9              & 35.3~$\pm$~0.7                & 30.0~$\pm$~0.3                        \\\relax
                [GY92] 30                       & 16 26 25.475          & -24 23 01.81            & 0.25~$\pm$~0.01~$\times$~0.12~$\pm$~0.01              & \enspace28~$\pm$~1              & 105.0~$\pm$~0.5               & 80.2~$\pm$~0.2                        \\
                WL 12                   & 16 26 44.203          & -24 34 48.86           & 0.14~$\pm$~0.01~$\times$~0.11~$\pm$~0.01              & \enspace60~$\pm$~11             & 160.4~$\pm$~1.0               & 142.4~$\pm$~0.5                       \\
                LFAM 23 $^{c}$  & 16 26 59.166          & -24 34 59.07          & 0.09~$\pm$~0.02~$\times$~0.05~$\pm$~0.04                & \enspace38~$\pm$~45           & 17.6~$\pm$~0.3          & 16.9~$\pm$~0.2                        \\\relax
                [GY92] 197              & 16 27 05.252          & -24 36 30.14           & 0.44~$\pm$~0.01~$\times$~0.13~$\pm$~0.01              & 169~$\pm$~1                     & 128.3~$\pm$~0.6               & 71.1~$\pm$~0.2                        \\ 
                Elias 29                        & 16 27 09.416          & -24 37 19.20            & 0.17~$\pm$~0.02~$\times$~0.16~$\pm$~0.02              & \enspace61~$\pm$~90             & 41.2~$\pm$~0.6                & 34.1~$\pm$~0.3                        \\
                IRS 43 VLA 1            & 16 27 26.908          & -24 40 50.67           & 0.21~$\pm$~0.03~$\times$~0.13~$\pm$~0.02              & \enspace96~$\pm$~17             & 41.6~$\pm$~1.9                & 32.1~$\pm$~0.9                        \\
                IRS 43 VLA 2            & 16 27 26.917          & -24 40 51.18           & 0.36~$\pm$~0.09~$\times$~0.2~$\pm$~0.1                        & 150~$\pm$~37                    & 7.5~$\pm$~1.0         & 3.7~$\pm$~0.4                 \\
                IRS 44                  & 16 27 27.988          & -24 39 33.93           & 0.24~$\pm$~0.03~$\times$~0.18~$\pm$~0.03              & 124~$\pm$~19                    & 38.6~$\pm$~1.2                & 29.1~$\pm$~0.6                        \\ 
                IRAS 16253-2429 & 16 28 21.620          & -24 36 24.17          & 0.23~$\pm$~0.02~$\times$~0.13~$\pm$~0.02                & 111~$\pm$~6                           & 39.8~$\pm$~0.7          & 31.9~$\pm$~0.4                        \\
                ISO-Oph 203             & 16 31 52.445          & -24 55 36.48           & 0.14~$\pm$~0.03~$\times$~0.05~$\pm$~0.03              & 107~$\pm$~18                    & 11.3~$\pm$~0.3                & 10.6~$\pm$~0.2                        \\
                IRS 67 A                        & 16 32 00.989          & -24 56 42.78            & \quad0.5~$\pm$~0.1~$\times$~0.3~$\pm$~0.2             & \enspace46~$\pm$~17             & 35.7~$\pm$~7.0                & 16.2~$\pm$~2.3                        \\
                IRS 67 B                        & 16 32 00.978          & -24 56 43.44            & 0.22~$\pm$~0.02~$\times$~0.14~$\pm$~0.01              & \enspace88~$\pm$~9              & 155.5~$\pm$~4.0               & 125.1~$\pm$~2.0                       \\      \hline
        \end{tabular}
        \tablefoot{$^{(a)}$ Deconvolved size (FWHM). $^{(b)}$ Brightness at the continuum peak. $^{(c)}$ Class~II T Tauri source located in the same field as IRAS 16238-2428.}
\end{table*}

The calibration and imaging were done in CASA\footnote{\tt http://casa.nrao.edu/} \citep{McMullin2007}: the complex gains were calibrated through observations of the quasars J1517-2422 and J1625-2527, passband calibration was based on J1924-2914, and flux calibration was based on Titan. A robust weighting with the Briggs parameter set to 0.5 was applied to the visibilities, and the resulting dataset has a typical beam size of 0\farcs43~$\times$~0\farcs32 ($\sim$60~$\times$~40~au), a continuum \textit{rms} level of 0.3~mJy~beam$^{-1}$ , and a spectral \textit{rms} level of 13 and 9~mJy~beam$^{-1}$ per 0.11 and 0.22~km~s$^{-1}$, respectively.

\section{Results}

\subsection{Continuum}

The sources that show continuum emission at 0.87~mm are listed in Table~\ref{table:coordinates}, with the sub-millimetre coordinates, deconvolved sizes, and fluxes calculated by fitting two-dimensional (2D) Gaussians in the image plane. Two of the sources, IRS 43 \citep{Girart2000} and IRS 67 \citep{McClure2010}, are known binary systems, and their components are separated by $\sim$70~au (0$\farcs$53) and $\sim$90~au (0$\farcs$68) for IRS 43 and IRS 67, respectively. The sources associated with a single continuum peak have diameters from $\sim$20~au (0$\farcs$14 for ISO-Oph 203) to $\sim$60~au (0$\farcs$44 for [GY92] 197) and fluxes from 11.3~mJy (ISO-Oph 203) to 160.4~mJy (WL 12). The deconvolved sizes show that most of the sources are marginally resolved, with the exception of LFAM 23, IRS 43 VLA 2, and IRS 67 A, which are unresolved.  

J162614.6 and IRAS 16238-2428 show neither continuum nor molecular line emission; the upper limit is 0.9~mJy~beam$^{-1}$ (3$\sigma$) for the continuum emission. J162614.6 shows faint emission at 24~$\mu$m and a low concentrated SCUBA\footnote{The Submillimeter Common-User Bolometer Array (SCUBA) on the James Clerk Maxwell Telescope (JCMT)} core \citep{Jorgensen2008}, suggesting that this source is more likely related with a prestellar core than with a protostar. For IRAS 16238-2428, the separation between its \textit{Spitzer} position and SCUBA core is 38$\farcs$6 \citep{Jorgensen2008}; the SCUBA core lies beyond the ALMA primary beam ($\sim$18$\arcsec$). However, the continuum emission detected in the same field of view as IRAS 16238-2428 may correspond to a T Tauri source (LFAM 23) that is located $\sim$4$\arcsec$ offset from the \textit{Spitzer} position that is associated with IRAS 16238-2428. Table~\ref{table:coordinates} presents results from a 2D Gaussian fit for LFAM 23, revealing that it is a point source. However, this source does not show molecular line emission and is therefore not included in the discussion in this paper.

Figure~\ref{fig:Continuum} shows the continuum emission of the remaining ten sources. For the binary systems, the continuum also traces the circumbinary material, which is remarkably extended for IRS 67 \citep[$\sim$550~au; for a detailed analysis of IRS 67, see][]{Artur2018}. It is worth mentioning that IRS 43 VLA 1 and IRS 67 B are brighter than IRS 43 VLA 2 and IRS 67 A at 0.87~mm, while the opposite situation is observed at infrared wavelengths \citep{Haisch2002, McClure2010}.

\subsection{Molecular lines}

The molecular transitions trace different components and are related with different morphologies. Some of them are detected towards the source position, while others peak offset from the source, and the emission may present compact or extended structures. Table~\ref{table:molecules} summarises the regions where the individual molecular transitions are detected, specifying if the emission is compact or extended (with respect to the extent of the continuum emission), and the source velocity (\textit{V$_\mathrm{source}$}). The latter is estimated by visual inspection of C$^{17}$O emission, or SO$_{2}$ emission when C$^{17}$O is not detected. For sources where no molecular line emission is detected, \textit{V$_\mathrm{source}$} is taken from the literature (see Table~\ref{table:molecules}). The two C$_{2}$H lines listed in Table~\ref{table:observations} are both hyperfine transitions, and the C$_{2}$H emission arises from a blended doublet, C$_{2}$H \textit{N}=4$-$3, \textit{J}=9/2$-$7/2, \textit{F}=5$-$4 plus C$_{2}$H \textit{N}=4$-$3, \textit{J}=9/2$-$7/2, \textit{F}=4$-$3. Emission of at least one molecular transition is detected towards eight of the ten sources. WL 12 (the brightest continuum source) and ISO-Oph 203 (the weakest continuum source) are the only sources where no line emission is detected within the spectral setting shown in Table~\ref{table:observations}.

\subsubsection{Optically thin tracers: C$^{17}$O, H$^{13}$CO$^{+}$ , and C$^{34}$S}

\begin{figure*}[hbtp]
   \centering
         \includegraphics[width=.7\textwidth]{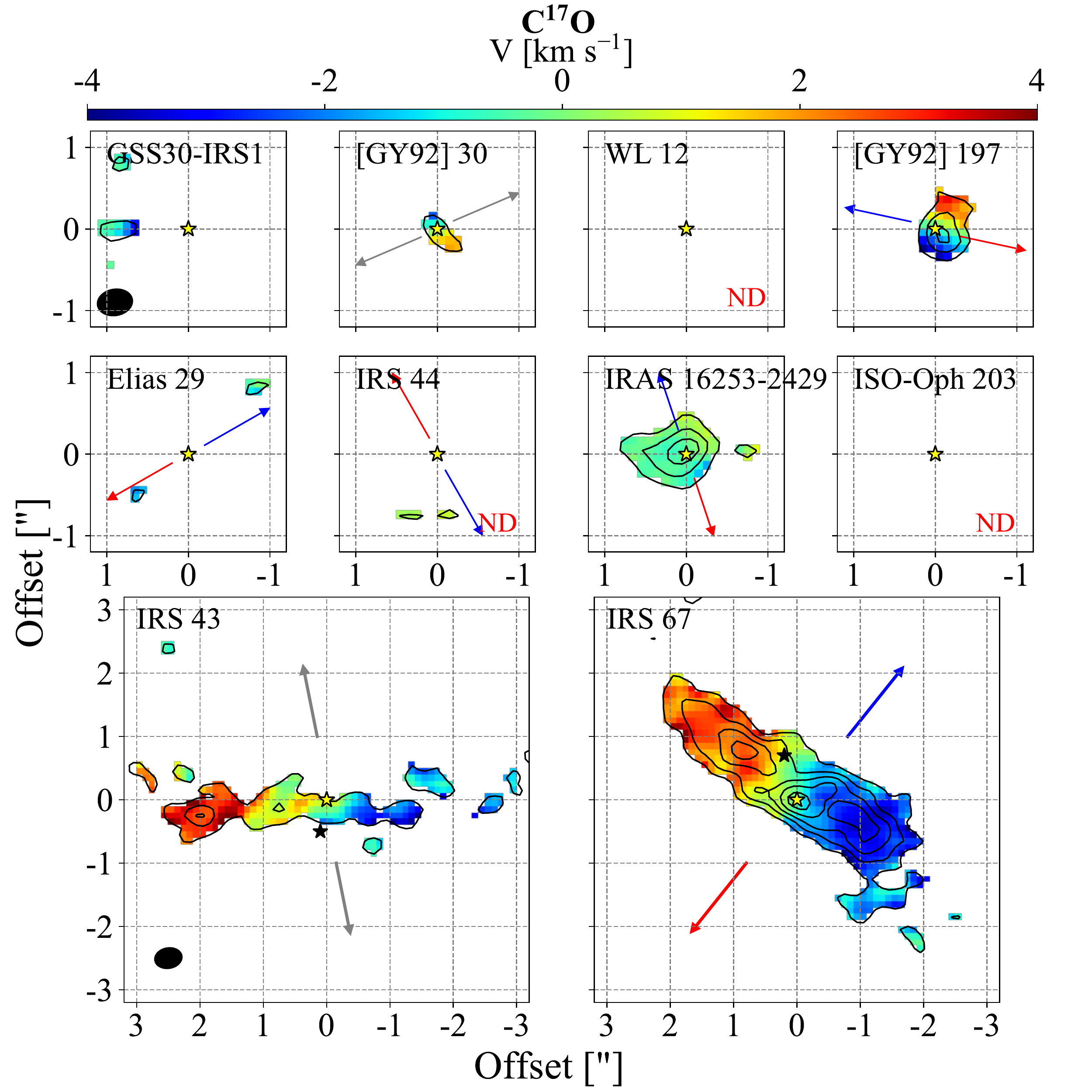}
      \caption[]{\label{fig:C17O}
       Moment 0 (black contours) and moment 1 (colour scale) maps for C$^{17}$O \textit{J}=3-2 above 5$\sigma$ and integrated over a velocity range of 8~km~s$^{-1}$. The contours start at 5$\sigma$ and follow a step of 4$\sigma$ ($\sigma$~=~4~mJy~beam$^{-1}$~km~s$^{-1}$). The yellow star indicates the position from a 2D Gaussian fit, and the black stars in the lower panels represent the position of the binary components. The arrows show the outflow direction from the literature, where the blue and red represent blue- and red-shifted emission, respectively, and the grey arrow represents infrared observations. The typical synthesised beam is shown by the black filled ellipse in the upper  and lower left panels. The ND label marks a non-detection.
      }
\end{figure*}

\begin{figure*}[hbtp]
   \centering
      \includegraphics[width=.61\textwidth]{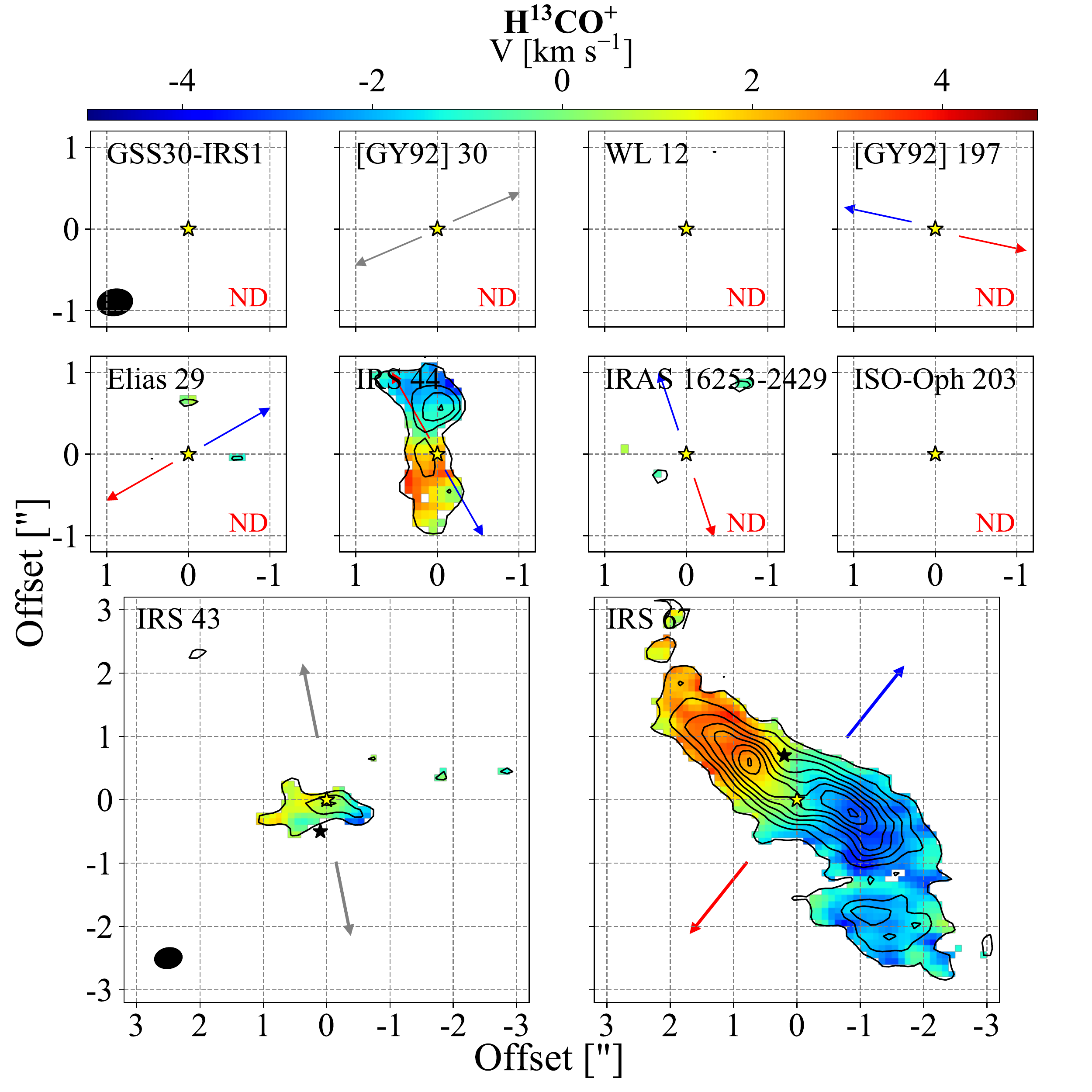}
      \caption[]{\label{fig:H13CO+}
      Same as Fig.~\ref{fig:C17O} for H$^{13}$CO$^{+}$ \textit{J}=4$-$3, integrated over a velocity range of 10~km~s$^{-1}$. The contours start at 5$\sigma$ and follow a step of 3$\sigma,$ with the exception of IRS 67, which follows a step of 6$\sigma$.
      }
\end{figure*}

\begin{figure*}[hbtp]
   \centering
      \includegraphics[width=.61\textwidth]{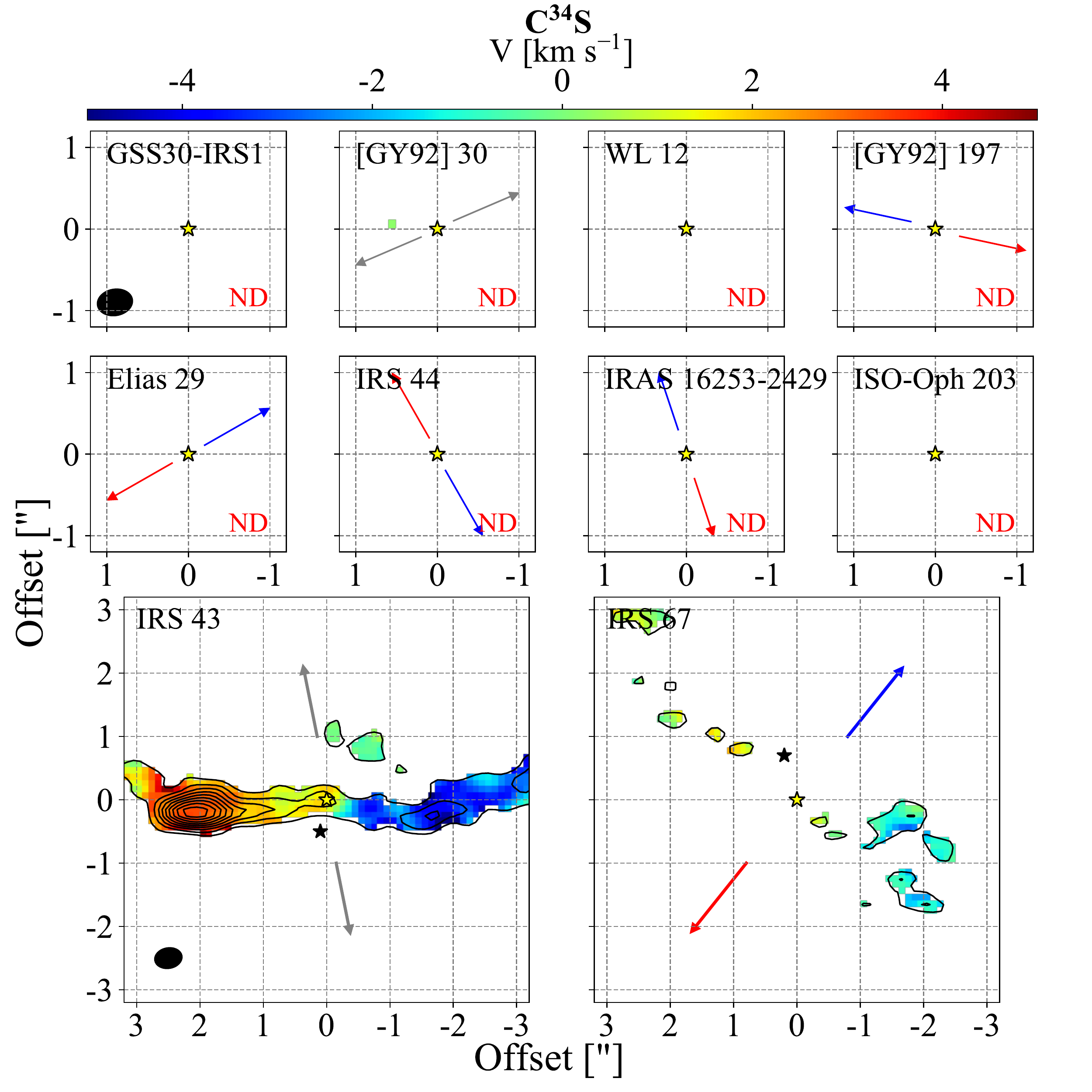}
      \caption[]{\label{fig:C34S}
      Same as Fig.~\ref{fig:C17O} for C$^{34}$S \textit{J}=7$-$6, integrated over a velocity range of 10~km~s$^{-1}$. The contours start at 5$\sigma$ and follow a step of 3$\sigma,$ with the exception of IRS 43, which follows a step of 6$\sigma$.
      }
\end{figure*}

C$^{17}$O, H$^{13}$CO$^{+}$ , and C$^{34}$S are less abundant isotopologues, and they are expected to be optically thin tracers associated with the disc kinematics. For example, a Keplerian disc was detected towards the Class~0/I source R CrA in similar C$^{17}$O observations with ALMA \citep{Lindberg2014}. 

Figure~\ref{fig:C17O} shows moment 0 and 1 maps for C$^{17}$O, indicating the outflow direction, if available, from the literature \citep{Bontemps1996, Allen2002, vanderMarel2013, Zhang2013, White2015, Yen2017}. Five of the sources, [GY92] 30, [GY92] 197, IRAS 16253-2429, IRS 43, and IRS 67, show on-source C$^{17}$O emission, and with the exception of IRAS 16253-2429, they are associated with a rotational profile perpendicular to the outflow direction, consistent with a disc-like structure (Sect. 4.2). In addition, the binary systems (IRS 43 and IRS 67) show C$^{17}$O emission that is much more extended than towards the other sources: 6$\farcs$1 diameter for IRS 43, 5$\farcs$4 for IRS 67, 1$\farcs$3 for IRAS 16253-2429, 0$\farcs$9 for [GY92] 197, and 0$\farcs$7 for [GY92] 30. The emission towards IRAS 16253-2429 is slightly elongated in the direction perpendicular to the outflow axis and is associated with low velocities (< $\pm$2~km~s$^{-1}$). This low-velocity component may be tracing infalling material from the inner envelope (see detailed discussion in Sect. 4.2). 

Figures~\ref{fig:H13CO+} and~\ref{fig:C34S} show images of the H$^{13}$CO$^{+}$ and C$^{34}$S emission. H$^{13}$CO$^{+}$ is detected towards only three of the ten sources: IRS 44 and the two binary systems, IRS 43 and IRS 67. IRS 44 shows a velocity gradient with two components, one consistent with the outflow direction, and the another rotated by $\sim$45$\degr$. The peak of emission (moment 0), however, is related with more quiescent material. The more massive of the binary systems (IRS 67) shows more extended emission, and its velocity profile is perpendicular to the outflow direction, while the emission towards IRS 43 is concentrated in the inner regions ($\leq$~1$\arcsec$) and a tentative velocity gradient is detected. In addition, C$^{34}$S is solely seen towards the binary systems. IRS 43 shows a velocity gradient perpendicular to the infrared jet direction and the emission peaks spatially offset from the system ($\sim$2$\arcsec$), where the red-shifted emission stands out. A different situation is observed for IRS 67: it has no clear velocity gradient and shows isolated peaks of emission that are spatially offset from the system.

\begin{table*}[t]
\caption{Regions where the individual molecular transitions are detected above 3$\sigma$ towards each source and the estimated source velocity (\textit{V$_\mathrm{source}$}).}
\label{table:molecules}
\centering
\begin{tabular}{l l l l l l l l}
        \hline\hline
                Source &  \multicolumn{6}{c}{Molecular transitions} & \textit{V$_\mathrm{source}$} $^{a}$\\
                 & C$^{17}$O & H$^{13}$CO$^{+}$ & C$^{34}$S & SO$_{2}$ & CH$_{3}$OH  & C$_{2}$H & [km s$^{-1}$] \\
        \hline
                GSS30-IRS1              & Offset                                & \multicolumn{1}{c}{-}   & \multicolumn{1}{c}{-} & On source [C]         & \multicolumn{1}{c}{-}   & \multicolumn{1}{c}{-}         & 3.4                   \\\relax
                [GY92] 30               & On source [C]         & \multicolumn{1}{c}{-}         & \multicolumn{1}{c}{-}         & \multicolumn{1}{c}{-} & Offset                                & Offset                          & 3.1                   \\
                WL 12                   & \multicolumn{1}{c}{-}         & \multicolumn{1}{c}{-}   & \multicolumn{1}{c}{-}         & \multicolumn{1}{c}{-} & \multicolumn{1}{c}{-}   & \multicolumn{1}{c}{-}         & 4.0 $^{b}$            \\\relax
                [GY92] 197              & On source                     & \multicolumn{1}{c}{-}   & \multicolumn{1}{c}{-}         & \multicolumn{1}{c}{-} & \multicolumn{1}{c}{-}   & On source [C]                 & 3.1                   \\
                Elias 29                        & Offset                                & \multicolumn{1}{c}{-}   & \multicolumn{1}{c}{-}         & On source [C]         & \multicolumn{1}{c}{-}   & \multicolumn{1}{c}{-}         & 3.6                   \\
                IRS 43                  & On source [E]         & On source                         & On source [E]         & On source B [C]               & \multicolumn{1}{c}{-}   & Offset                                & 4.0                   \\
                IRS 44                  & \multicolumn{1}{c}{-} & On source [E]             & \multicolumn{1}{c}{-}         & On source [C]         & \multicolumn{1}{c}{-}   & \multicolumn{1}{c}{-} & 2.3                   \\
                IRAS 16253-2429         & On source [E]         & \multicolumn{1}{c}{-} & \multicolumn{1}{c}{-}   & \multicolumn{1}{c}{-} & \multicolumn{1}{c}{-} & On source [E]           & 4.0                   \\
                ISO-Oph 203             & \multicolumn{1}{c}{-}         & \multicolumn{1}{c}{-}   & \multicolumn{1}{c}{-}         & \multicolumn{1}{c}{-} & \multicolumn{1}{c}{-}   & \multicolumn{1}{c}{-} & 4.5 $^{b}$            \\
                IRS 67                  & On source [E]         & On source [E]             & Offset                                & On source B [C]               & \multicolumn{1}{c}{-}   & On source [E]         & 4.2                   \\
        \hline
\end{tabular}
\tablefoot{$^{(a)}$ Estimated from visual inspection of the C$^{17}$O spectrum, or SO$_{2}$ when C$^{17}$O is not detected. $^{(b)}$ Values taken from \cite{Lindberg2017}. The [C] and [E] labels refer to compact or extended emission (with respect to the extension of the continuum emission).}
\end{table*}

\subsubsection{Warm chemistry tracers: CH$_{3}$OH and SO$_{2}$}

\begin{figure}[hbtp]
   \centering
      \includegraphics[width=.49\textwidth]{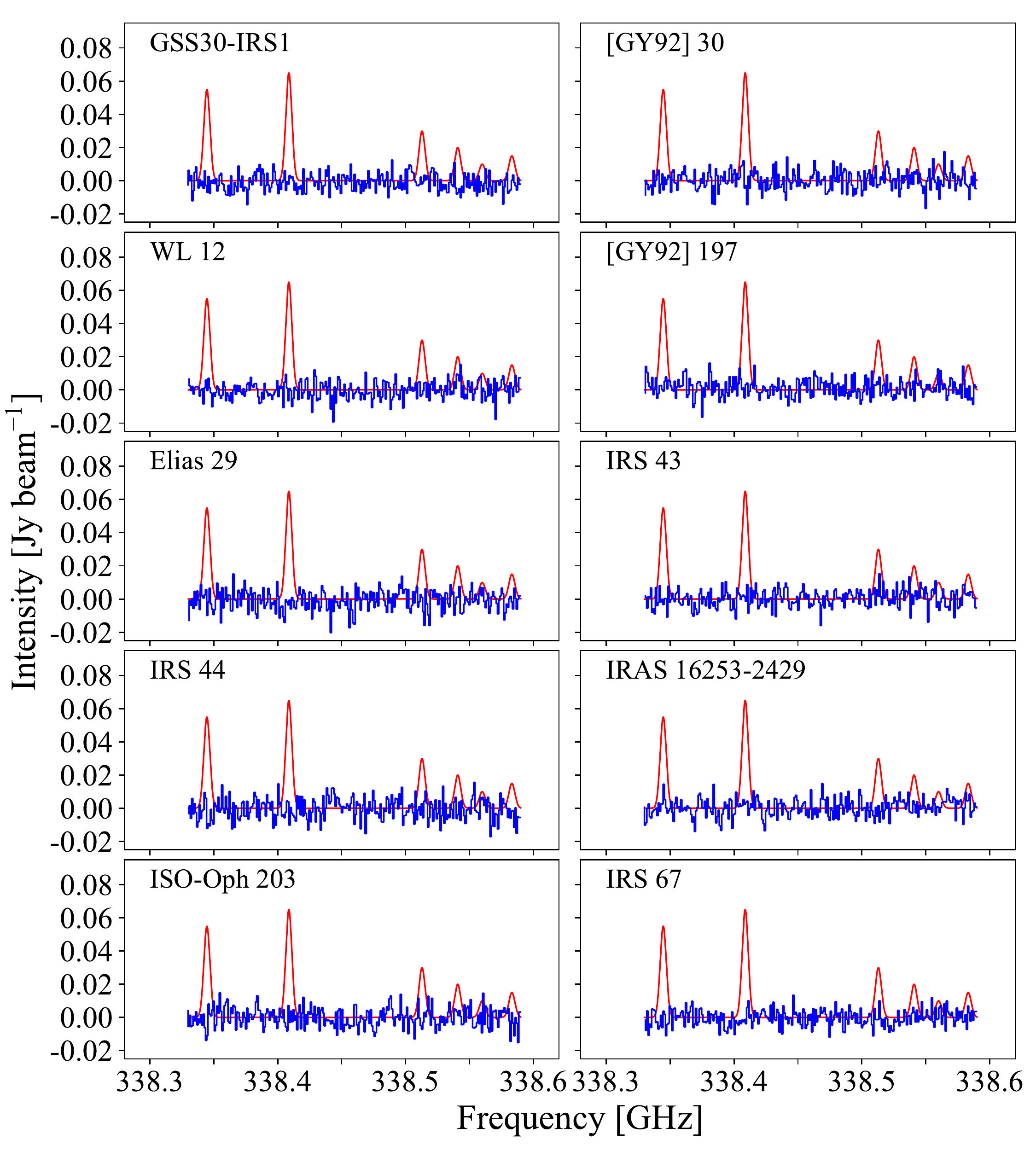}
      \caption[]{\label{fig:CH3OH_spectra}
      Observed CH$_{3}$OH spectra \textit{(blue)} towards the 10 sources that show continuum emission, superimposed on the predicted CH$_{3}$OH spectrum \textit{(red)} for a column density of 2.5~$\times$~10$^{15}$~cm$^{-2}$, $\sim$4 orders of magnitude below the value from the Class~0 source IRAS 16293-2422 \citep{Jorgensen2016}.
      }
\end{figure}

\begin{figure*}[hbtp]
   \centering
      \includegraphics[width=.61\textwidth]{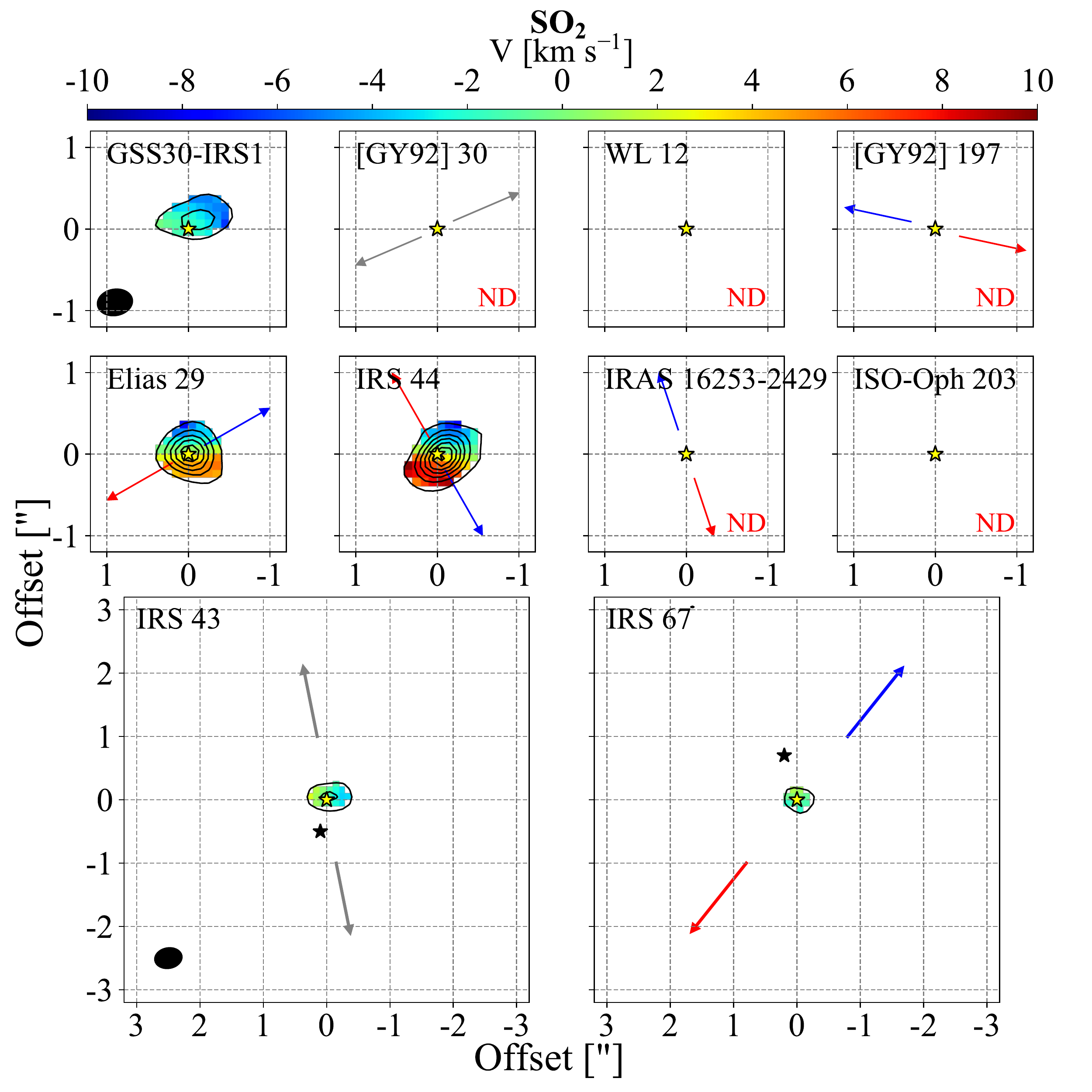}
      \caption[]{\label{fig:SO2}
      Same as Fig.~\ref{fig:C17O} for SO$_{2}$ \textit{J}$_\mathrm{K_{A}K_{C}}$=18$_{4,14}$$-$18$_{3,15}$, integrated over a velocity range of 20~km~s$^{-1}$. The contours start at 5$\sigma$ and follow a step of 20$\sigma$.
      }
\end{figure*}

\begin{figure*}[hbtp]
   \centering
      \includegraphics[width=.61\textwidth]{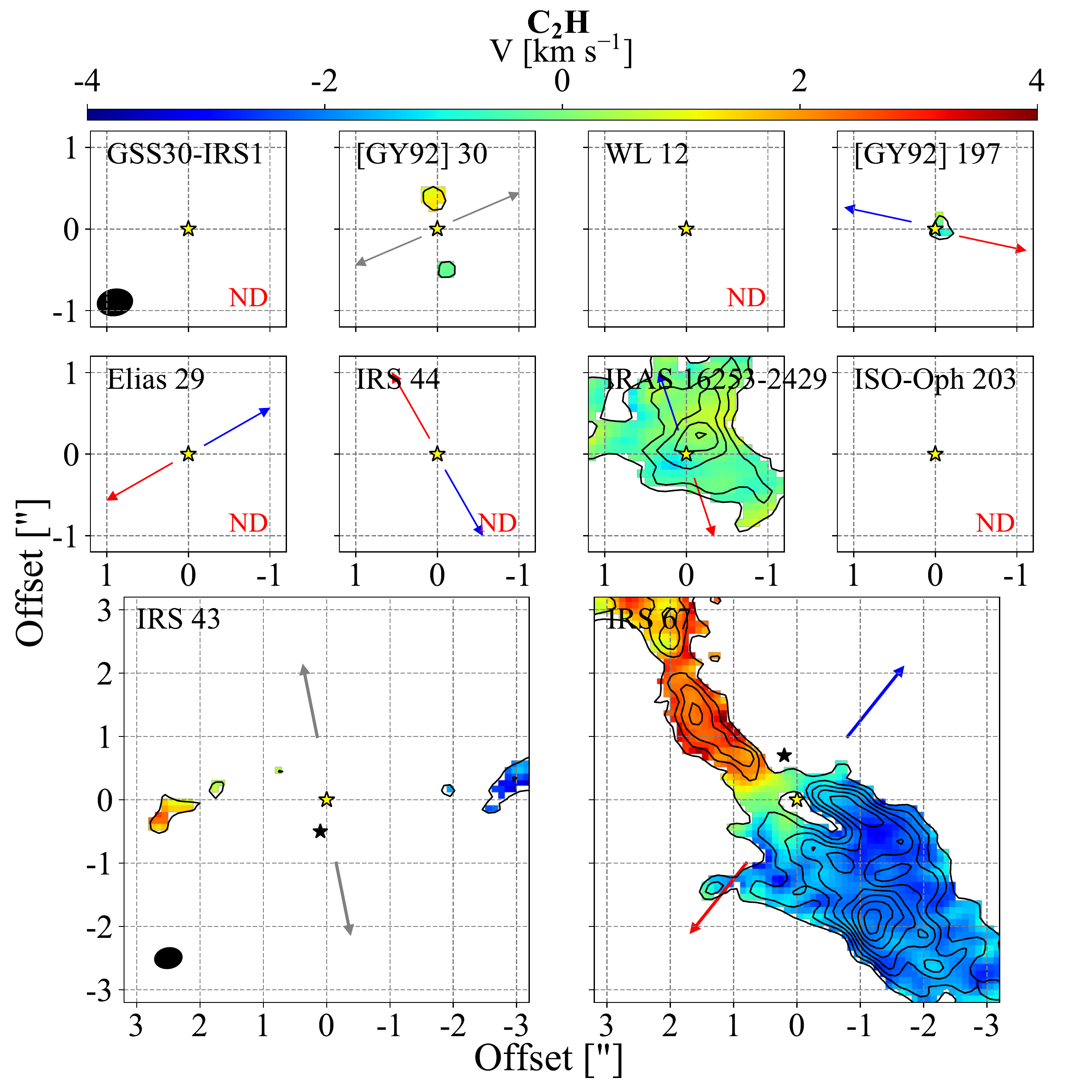}
      \caption[]{\label{fig:C2H}
      Same as Fig.~\ref{fig:C17O} for the doublet C$_{2}$H \textit{N}=4$-$3, \textit{J}=9/2$-$7/2. 
      }
\end{figure*}

CH$_{3}$OH forms exclusively on ice-covered dust grain surfaces because gas-phase reactions produce negligible CH$_{3}$OH abundances \citep{Garrod2006, Chuang2016, Walsh2016}. When the temperature reaches $\sim$90~K, CH$_{3}$OH thermally desorbes from the grain mantles, and its gas-phase abundance is enhanced close to the protostar \citep{Brown2007}. Therefore, CH$_{3}$OH transitions associated with high rotational levels, such as \textit{J}=7$_{k}$$-$6$_{k}$, are expected to trace the warm gas close to the protostar. 

The observed CH$_{3}$OH \textit{J}=7$_{k}$$-$6$_{k}$ branch includes transitions with \textit{E$_\mathrm{u}$} from 65 to 376~K, but no significant emission is detected towards the continuum position from a 2D Gaussian fit of any source (see Table~\ref{table:coordinates}). This is highlighted in Fig.~\ref{fig:CH3OH_spectra}, where the observed and predicted spectrum are plotted together. The predicted spectrum is obtained using the statistical equilibrium radiative transfer code \citep[RADEX; ][]{vanderTak2007} and assuming local thermodynamic equilibrium (LTE), a CH$_{3}$OH column density of 2.5~$\times$~10$^{15}$~cm$^{-2}$ \citep[$\sim$4 orders of magnitude below the CH$_{3}$OH column density measured towards the Class~0 source IRAS 16293-2422; ][]{Jorgensen2016}, and a kinetic temperature (\textit{T$_\mathrm{kin}$}) of 100~K. Clearly, no CH$_{3}$OH lines are seen even at this level. 

Nevertheless,  [GY92] 30 is a peculiar case (the least luminous source of the sample, \textit{L$_\mathrm{bol}$} = 0.12 \textit{L$_\mathrm{\odot}$}, and the source associated with the more massive envelope, \textit{M$_\mathrm{env}$} = 0.27~M$_{\odot}$), where two CH$_{3}$OH transitions are detected offset from the source, beyond the 25$\sigma$ continuum contour (Figs.~\ref{fig:CH3OH} and \ref{fig:CH3OH_bis} in the Appendix). These transitions are associated with the lowest \textit{E$_\mathrm{u}$} levels (65 and 70~K) and the emission is related with low velocities of between $-$0.5 and 0.5~km~s$^{-1}$ from the source velocity. Thus, the CH$_{3}$OH emission towards [GY92] 30 is likely related with extended envelope material. In addition, there is no clear association between the CH$_{3}$OH emission and the direction of the infrared jet.  

SO$_{2}$ is commonly associated with outflows and shocked regions \citep[e.g.][]{Jorgensen2004c, Persson2012, Podio2015, Tabone2017}. In addition, the highly excited rotational transitions (\textit{E$_\mathrm{u}$} $\sim$200~K) may be tracing warm shocked gas. The observed SO$_{2}$ transition, listed in Table~\ref{table:observations}, is associated with a high rotational level (\textit{E$_\mathrm{u}$} = 197~K) and is detected towards five of the sources, showing compact emission (Fig.~\ref{fig:SO2}). A rotational profile is seen for Elias 29 and IRS 44, associated with high velocities (up to $\pm$10~km~s$^{-1}$ with respect to the source velocity), and almost perpendicular to the outflow direction. Towards the binary systems, the SO$_{2}$ emission is relatively weak and detected around one of the sources, IRS 43 VLA1 and IRS 67 B; the sources correspond to the brightest components at 0.87~mm. In addition, GSS30-IRS1 shows only a blue-shifted component, without a clear velocity profile. The SO$_{2}$ emission towards Elias 29 and IRS 44 may be consistent with a disc-like structure, but the broad-line profile suggests that SO$_{2}$  traces a different component that is associated with warm shocked gas.

\subsubsection{Outer envelope tracer: C$_{2}$H}

The emission of C$_{2}$H has been associated with dense regions exposed to UV radiation \citep[e.g.][]{Nagy2015, Murillo2018}. Its emission is shown in Fig.~\ref{fig:C2H} and is detected towards five of the sources. Two of the them, [GY92] 197 and IRAS 16253-2429, show on-source emission related with low velocities, showing a compact morphology for [GY92] 197 and an extended structure for IRAS 16253-2429. Towards [GY92] 30, IRS 43, and IRS 67, there is no C$_{2}$H emission (above 5$\sigma$) at the positions of the sources, but a velocity gradient is seen for the circumbinary material, with a remarkable extended emission for IRS 67 ($\sim$9$\arcsec$). Because IRAS 16253-2429 has one of the most massive envelopes (\textit{M$_\mathrm{env}$} = 0.15~M$_{\odot}$) and circumbinary discs are usually related with a higher mass content and greater extension than circumstellar discs, C$_{2}$H likely traces more dense regions.

\section{Mass evolution}

The final mass of a protostar and the amount of material available in the disc to form planets depend on the mass evolution of the whole system (envelope, disc, protostar, and outflow). Determining when and how quickly the envelope dissipates and the disc grows in mass and size, the rate at which the protostar gains mass and the amount of material expelled through the outflows are all linked to each other and are crucial components in the mass evolution. In this section, we estimate the disc masses from the continuum emission (Sect.~4.1) and the stellar masses from molecular lines that show Keplerian profiles (Sect.~4.2). Later on, a comparison between envelope mass, disc mass, stellar mass, and bolometric luminosity is presented (Sect.~4.3). Finally, the mass accretion rate is estimated from a relationship between the stellar mass and the bolometric luminosity, and this is compared with the bolometric temperature.

\subsection{disc mass}

The disc masses (\textit{M$_\mathrm{disc}$}) were calculated from the continuum fluxes (\textit{F$_\mathrm{0.87mm}$}), listed in Table~\ref{table:coordinates}, and

\begin{equation} 
    M_\mathrm{disc} = \frac{S_{\nu} d^{2}}{\kappa_{\nu}B_{\nu}(T)} \ ,
    \label{eq:Eq2}
\end{equation}

\noindent where \textit{S$_{\nu}$} is the surface brightness, \textit{d} is the distance to the source, $\kappa$$_{\nu}$ is the dust opacity, and \textit{B$_{\nu}$(T)} is the Planck function for a single temperature. A distance of 139~$\pm$~6~pc \citep{Mamajek2008} and $\kappa$$_\mathrm{0.87mm}$ of 0.0175 cm$^{2}$~g$^{-1}$ \citep{Ossenkopf1994}, commonly used for dust in protostellar envelopes and young discs in the millimetre regime \citep[e.g.][]{Shirley2011}, was adopted for the calculations \citep{Artur2018}. A value of 15~K was adopted for the dust temperature (\textit{T$_\mathrm{dust}$}) following the analysis by \cite{Dunham2014b}. The calculated disc masses (gas + dust, assuming a gas-to-dust ratio of 100) are listed in Table~\ref{table:masses}, together with values from the literature. Errors in \textit{T$_\mathrm{dust}$} are not considered in Table~\ref{table:masses}, but the disc masses will decrease by a factor of $\sim$3 when a dust temperature of 30~K is assumed instead (found to be appropriate for Class~0 sources in the study by \citealt{Dunham2014b}). For WL 12, Elias 29, and IRS 43, disc masses are available from other studies (see Table~\ref{table:masses}). For Elias 29 and IRS 43, the values from this work are comparable with the literature, and the difference seen for WL 12 appears to be due to the choice of dust temperature (15 vs. 30~K). Assuming \textit{T$_\mathrm{dust}$}~=~30~K, a value of (10.3~$\pm$~2.4)~$\times$~10$^{-3}$~M$_{\odot}$ is found for \textit{M$_\mathrm{disc}$} associated with WL 12.

\begin{table}[t]
        \caption{Calculated disc masses for \textit{T$_\mathrm{dust}$}~=~15~K and values from the literature.}
        \label{table:masses}
        \centering
        \setlength{\tabcolsep}{11pt}    
        \begin{tabular}{l c r}
        \hline\hline
                \multicolumn{1}{c}{Source}      & \multicolumn{2}{c}{ \textit{M$_\mathrm{disc}$} [10$^{-3}$ M$_{\odot}$]}                \\
                                                                & This work                                     & Literature                                              \\
        \hline
                GSS30-IRS1              & 6.2~$\pm$~0.6                 &                               \\\relax
                [GY92] 30                       & 18.5~$\pm$~1.6                &                               \\
                WL 12                   & 28.3~$\pm$~2.4                & 11      $^{a}$          \\
                LFAM 23                 & 3.1~$\pm$~0.3                 &                               \\\relax
                [GY92] 197              & 22.6~$\pm$~2.0                &                               \\ 
                Elias 29                        & 7.3~$\pm$~0.6                 & 11 $^{a}$               \\
                                                &                                       & < 7 $^{b}$              \\
                IRS 43 VLA 1            & 7.3~$\pm$~0.7                 & 8 $^{a}$                        \\
                IRS 43 VLA 2            & 1.3~$\pm$~0.2                 &                               \\
                IRS 44                  & 6.8~$\pm$~0.6                 &                               \\ 
                IRAS 16253-2429 & 7.0~$\pm$~0.6                 &                               \\
                ISO-Oph 203             & 2.0~$\pm$~0.2                 &                               \\
                IRS 67 A                        & 6.3~$\pm$~1.3                 &                               \\      
                IRS 67 B                        & 27.4~$\pm$~2.5                &                               \\      
        \hline
        \end{tabular}
        \tablefoot{The errors in the second column do not include uncertainties in the assumed dust temperature (\textit{T$_\mathrm{dust}$} = 15~K), but the disc masses decrease by a factor of $\sim$3 when a dust temperature of 30~K is assumed. $^{(a)}$ From \cite{Jorgensen2009} at 1.1~mm, with \textit{T$_\mathrm{dust}$}~=~30~K. $^{(b)}$ From \cite{Lommen2008} at 1.1~mm, with \textit{T$_\mathrm{dust}$}~=~30~K.}
\end{table}

\subsection{Stellar mass}

\begin{figure*}[hbtp]
   \centering
      \includegraphics[width=.32\textwidth]{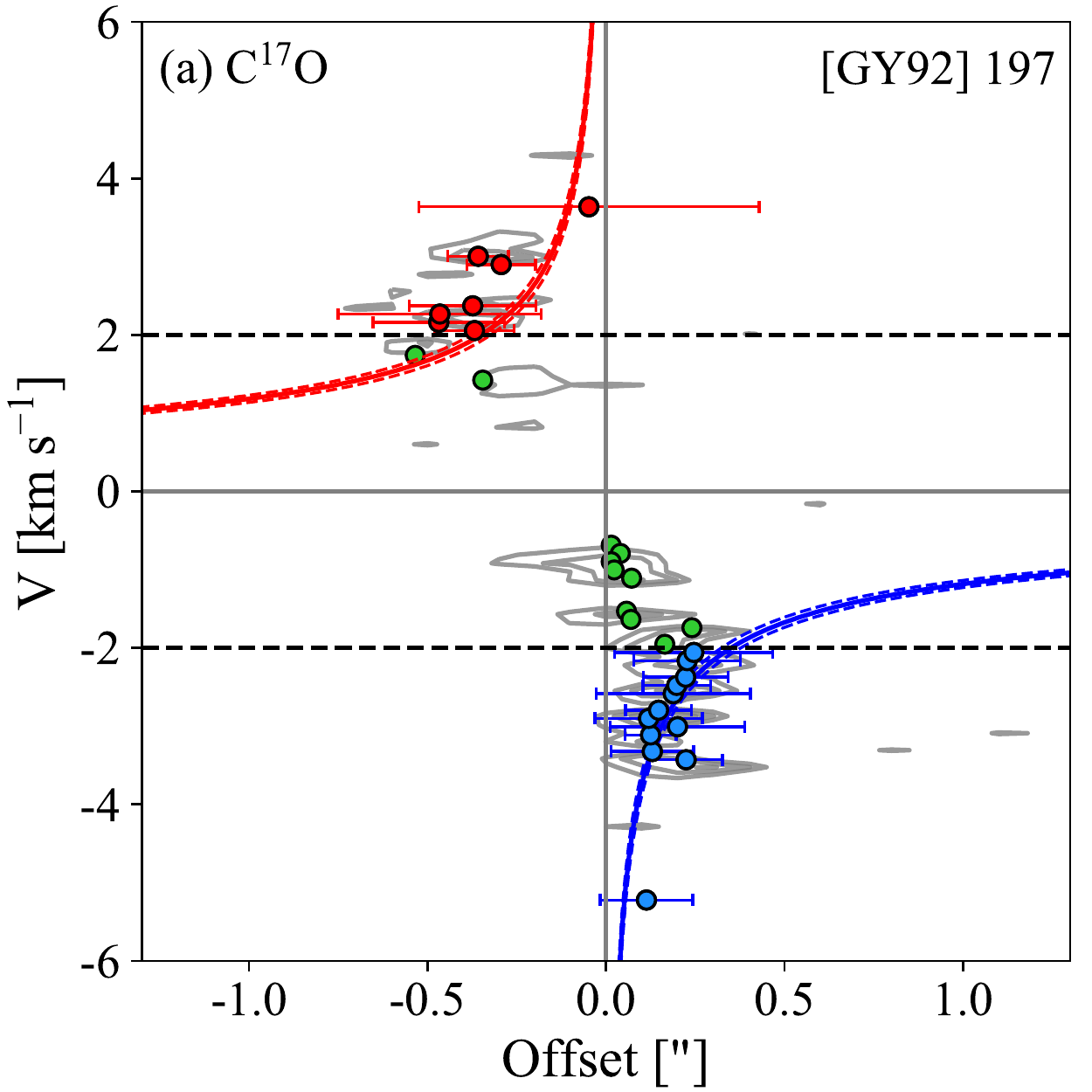}
      \includegraphics[width=.32\textwidth]{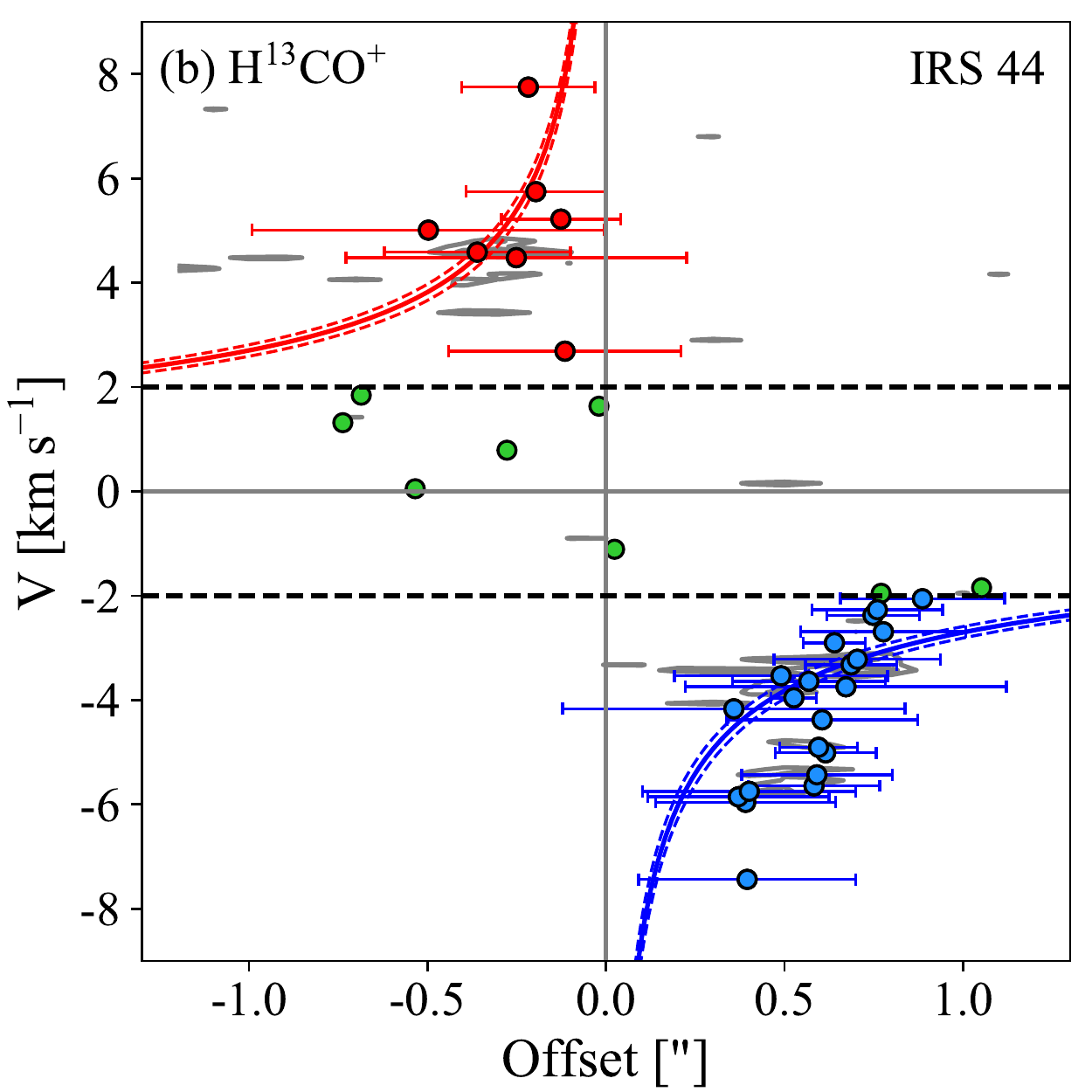}
      \includegraphics[width=.32\textwidth]{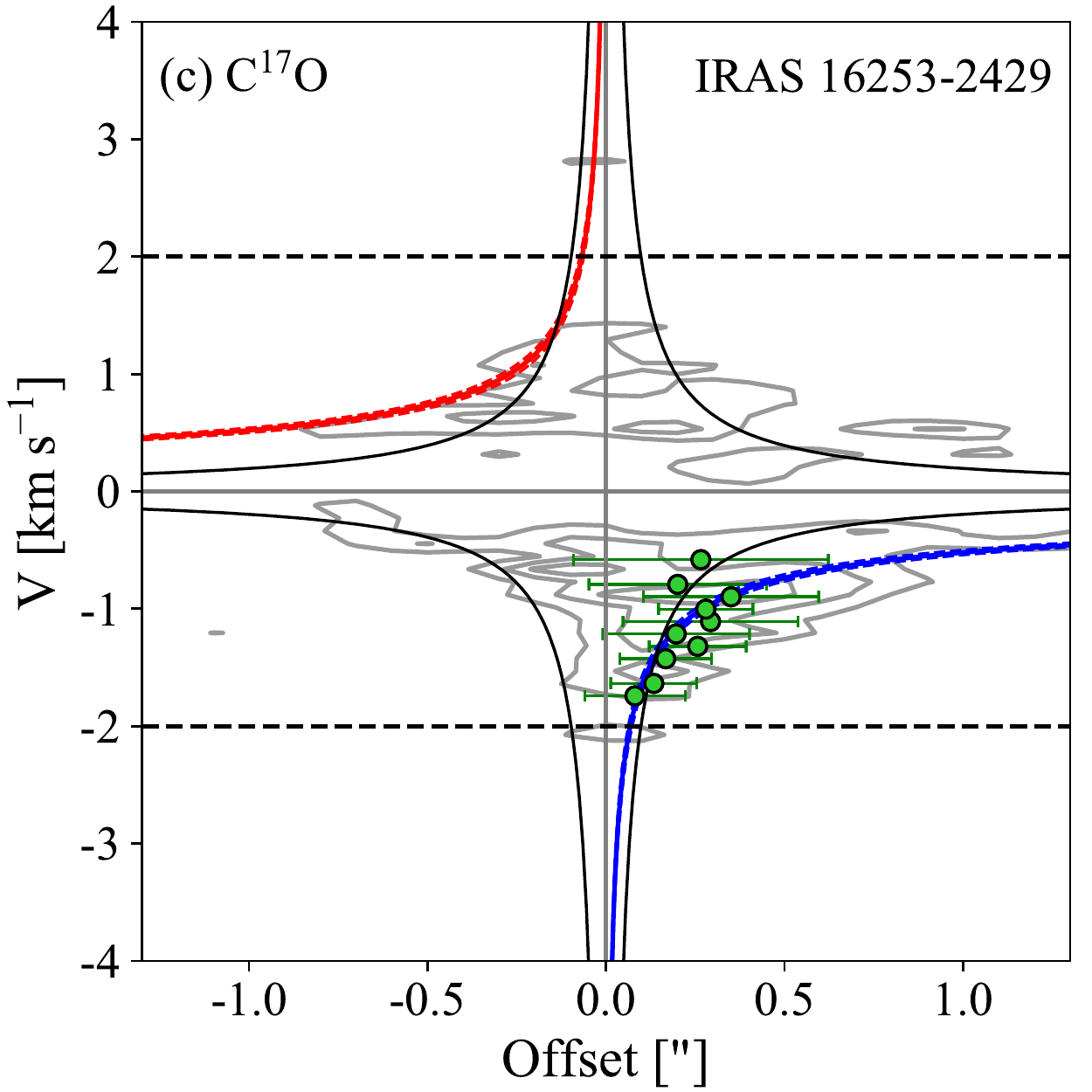}
      \caption[]{\label{fig:PV1}
      Position-velocity diagrams towards [GY92] 197, IRS 44 and IRAS 16253-2429. Blue and red dots represent blue- and red-shifted emission peaks above $\pm$2~km~s$^{-1}$, respectively, while green dots indicate velocities below $\pm$2~km~s$^{-1}$. Blue and red solid lines show the best fit for a Keplerian velocity profile with their respective errors, shown in dashed blue and red lines. The cut taken from the image data is shown in grey contours, ranging from 3$\sigma$ ($\sigma$ = 13~mJy~beam$^{-1}$) to the maximum value of each transition. Each adjacent contour represents an increment of 30$\%$ of the maximum value for panels \textit{(a)} and \textit{(c)}, and an increment of 50$\%$ for panel \textit{(b)}. The black dashed lines indicate the velocity above which the Keplerian profile was fitted. The solid black curves in panel \textit{(c)} show the best fit for an infalling velocity profile.
      }
\end{figure*}

\begin{figure*}[hbtp]
   \centering
      \includegraphics[width=.4\textwidth]{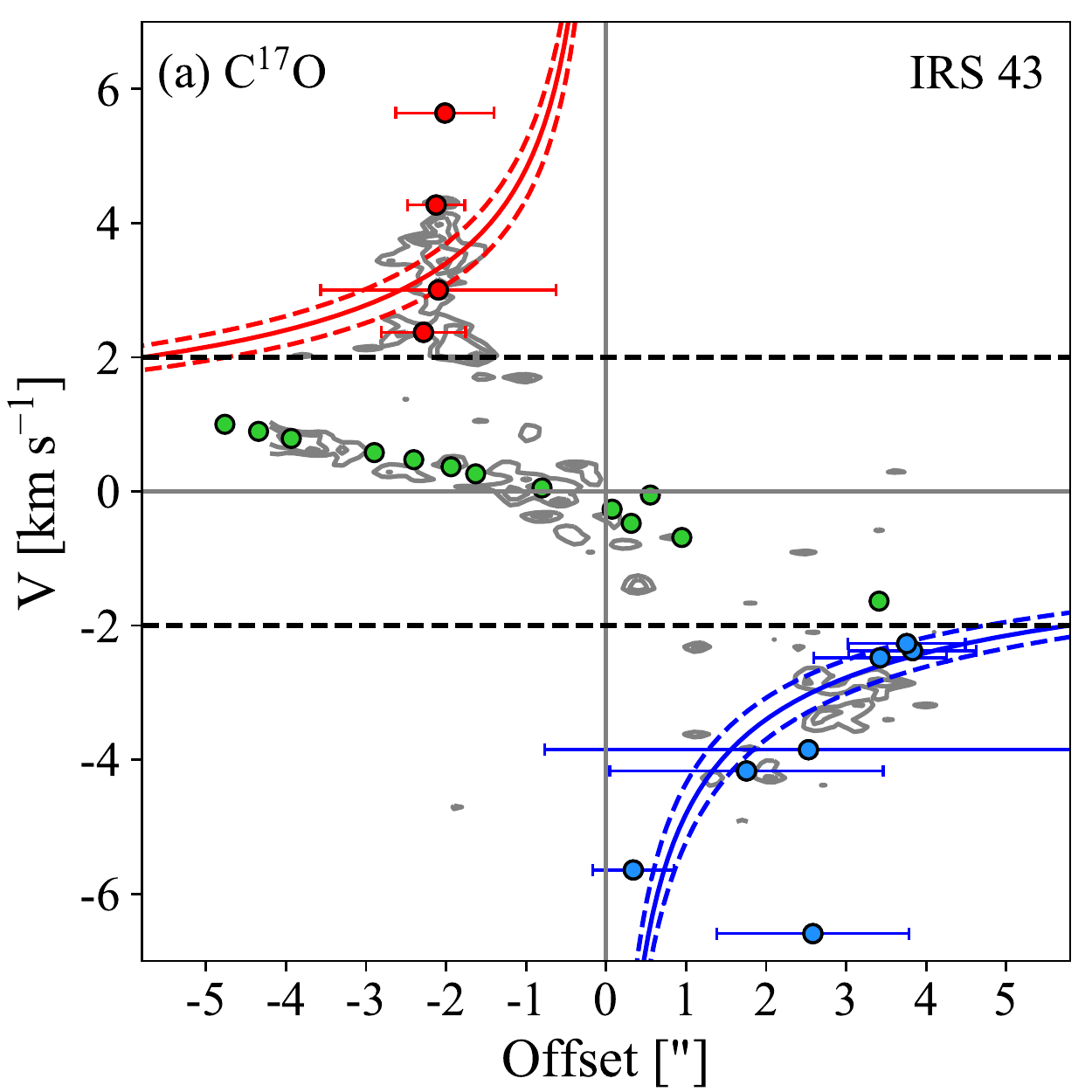}
      \includegraphics[width=.4\textwidth]{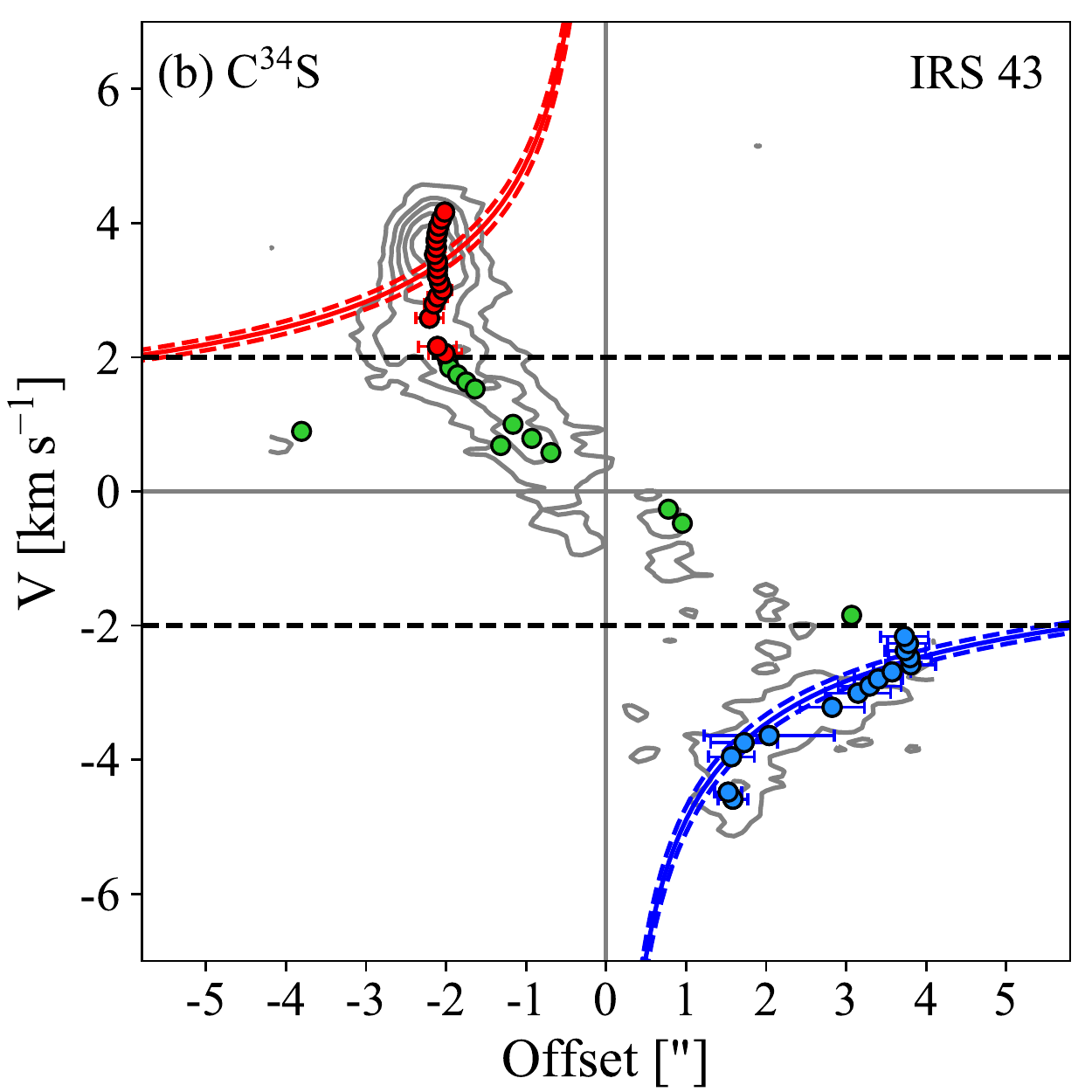}
      \caption[]{\label{fig:PV2}
      Position-velocity diagrams for C$^{17}$O and C$^{34}$S towards IRS 43. Blue and red dots represent blue- and red-shifted emission peaks above $\pm$2~km~s$^{-1}$, respectively, while green dots indicate velocities below $\pm$2~km~s$^{-1}$. Blue and red solid lines show the best fit for a Keplerian velocity profile with their respective errors, shown in dashed blue and red lines. The cut taken from the image data is shown in grey contours, ranging from 3$\sigma$ ($\sigma$ = 13~mJy~beam$^{-1}$) to the maximum value of each transition. Each adjacent contour represents an increment of 30 and 20$\%$ of the maximum value for C$^{17}$O and C$^{34}$S, respectively. The black dashed lines indicate the velocity above which the Keplerian profile was fitted.
      }
\end{figure*}

In order to investigate whether the optically thin tracers are associated with Keplerian motions and to estimate the stellar masses (\textit{M$_{\star}$}), position-velocity (PV) diagrams were created for the sources that show disc-like structures in C$^{17}$O, H$^{13}$CO$^{+}$ or C$^{34}$S. The peak emission for each channel was obtained through the CASA task \texttt{imfit,} and the offset position was calculated by projecting the peak emission onto the disc position angle (Table~\ref{table:coordinates}). Next, a Keplerian profile (\textit{$v$} $\propto$ \textit{r$^{-0.5}$}) was employed to fit the points with velocities above $\pm$2~km~s$^{-1}$. This velocity range was chosen in order to avoid the envelope contribution \citep[e.g.][]{vantHoff2018}. The PV diagrams and peak points are shown in Figs.~\ref{fig:PV1} and \ref{fig:PV2}, and the resulting Keplerian curve from the fit is overplotted. Figure~\ref{fig:PV1}a shows a robust Keplerian profile for [GY92] 197, while the PV diagram for IRS 44 is noisy (Fig.~\ref{fig:PV1}b), with an unclear Keplerian profile. IRAS 16253-2429 shows low-velocity emission ($\leq$2~km~s$^{-1}$) and both negative and positive velocities for the same distance to the source (see Fig.~\ref{fig:PV1}c). This is a characteristic feature of infalling material \citep[e.g.][]{Tobin2012}. Therefore, these points were fitted with an infalling profile under the conservation of angular momentum \citep[\textit{$v$}~$\propto$~\textit{r$^{-1}$}; e.g.][]{Lin1994, Harsono2014}. The resulting stellar masses are listed in Table~\ref{table:Fits}, adopting an inclination (\textit{i}) of 70$\degr$ from the plane of the sky. This value is adopted from the mean value of the inclination of the sample, calculated from the deconvolved continuum sizes (Table~\ref{table:coordinates}) and assuming a circular structure. In addition, 70$\degr$ is consistent with values from the literature: 70$\degr$ for IRS 43 \citep{Brinch2016} and 60$\degr$ for IRAS 16253-2429 \citep{Yen2017}. The stellar mass changes by less than 14$\%$ when the inclination varies with $\pm$10$\degr$.

IRS 43 shows extended emission in C$^{17}$O and C$^{34}$S, with PV diagrams and Keplerian fits shown in Fig.~\ref{fig:PV2} for both molecular transitions. The blue-shifted C$^{34}$S emission is well fitted with a Keplerian profile, but the red-shifted emission shows an enhancement around 2$\arcsec$ and a vertical velocity structure, from $\sim$2 to $\sim$4~km~s$^{-1}$. PV diagrams for IRS 67 from C$^{17}$O and H$^{13}$CO$^{+}$ emission are presented in \cite{Artur2018}.

Table~\ref{table:Fits} lists the stellar masses obtained from PV diagrams and values from the literature. For IRS 43, the stellar masses inferred from a Keplerian fit from C$^{17}$O and C$^{34}$S emission are a factor of 2 higher than the 1.80~M$_{\odot}$ from \cite{Brinch2016}. To be consistent within the sample, the stellar mass of 4.0~M$_{\odot}$ is used for the IRS 43 system hereafter. For IRAS 16253-2429, the stellar mass from \cite{Yen2017} is consistent with our value from the fit for the infalling gas.

\begin{table*}[t]
\caption{Stellar masses obtained from the velocity profile fits and values from the literature.}
\label{table:Fits}
\centering
\begin{tabular}{l c l l c c c l}
        \hline\hline
                Source                  &       & \multicolumn{2}{c}{This work}                                           & \multicolumn{4}{c}{Literature}                                                                                                \\
                                                &       & \textit{M$_{\star}$} [M$_{\odot}$] $^{a}$    & Molecule              &       &       & \textit{M$_{\star}$} [M$_{\odot}$]           & \multicolumn{1}{c}{Method}    \\
        \hline
                [GY92] 197              &       & 0.23 $\pm$ 0.02               & C$^{17}$O                               &       &       &                                               &                                                               \\
                Elias 29                        &       &                                       &                                               &       &       & 2.5 $\pm$ 0.6 $^{b}$            & HCO$^{+}$ \textit{J}=3$-$2 emission   \\
                IRS 43                  &       & 4.0 $\pm$ 0.3         & C$^{34}$S                               &       &       & 1.80 $\pm$ 0.42 $^{c}$          & HCN \textit{J}=3$-$2 emission         \\
                                                &       & 3.9 $\pm$ 0.7         & C$^{17}$O                               &       &       &                                                &                                                               \\
                IRS 44                  &       & 1.2 $\pm$ 0.1         & H$^{13}$CO$^{+}$                &       &       &                                               &                                                               \\
                IRAS 16253-2429 &       & 0.03 $\pm$ 0.02               & C$^{17}$O (infalling)           &       &       & 0.03 $\pm$ 0.01 $^{d}$                & C$^{18}$O \textit{J}=2$-$1 emission     \\
                IRS 67 $^{e}$           &       & 2.2 $\pm$ 0.2                 & C$^{17}$O                               &       &       &                                                &                                                               \\
        \hline
\end{tabular}
\tablefoot{$^{(a)}$ The stellar masses were calculated assuming an inclination of 70$\degr$. $^{(b)}$ From \cite{Lommen2008}. $^{(c)}$ From \cite{Brinch2016}, for the binary system. $^{(d)}$ From \cite{Yen2017}. $^{(e)}$ A more detailed analysis of IRS 67 is presented in \cite{Artur2018}.}
\end{table*}

\subsection{Mass evolution}

\begin{figure*}[hbtp]
   \centering
      \includegraphics[width=.95\textwidth]{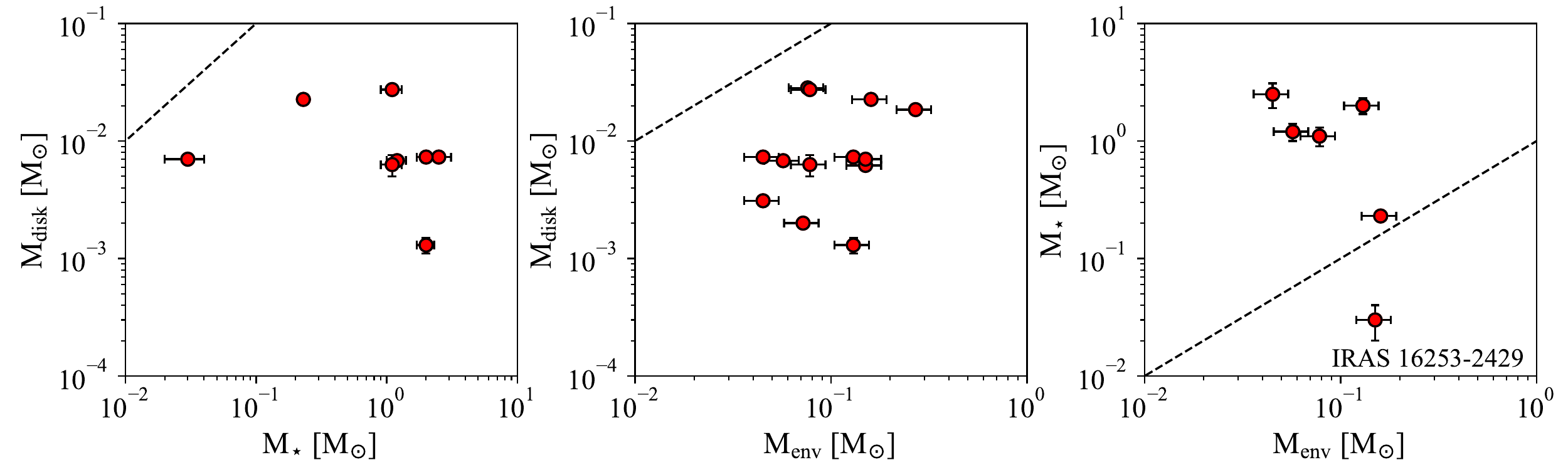}
      \caption[]{\label{fig:M_DES}
      Comparison between disc mass and stellar mass (\textit{left}), disc mass and envelope mass (\textit{centre}), and stellar mass and envelope mass (\textit{right}). The dashed black line is for \textit{M$_{i}$} = \textit{M$_{j}$}.
      }
\end{figure*}

\begin{figure*}[hbtp]
   \centering
      \includegraphics[width=.9\textwidth]{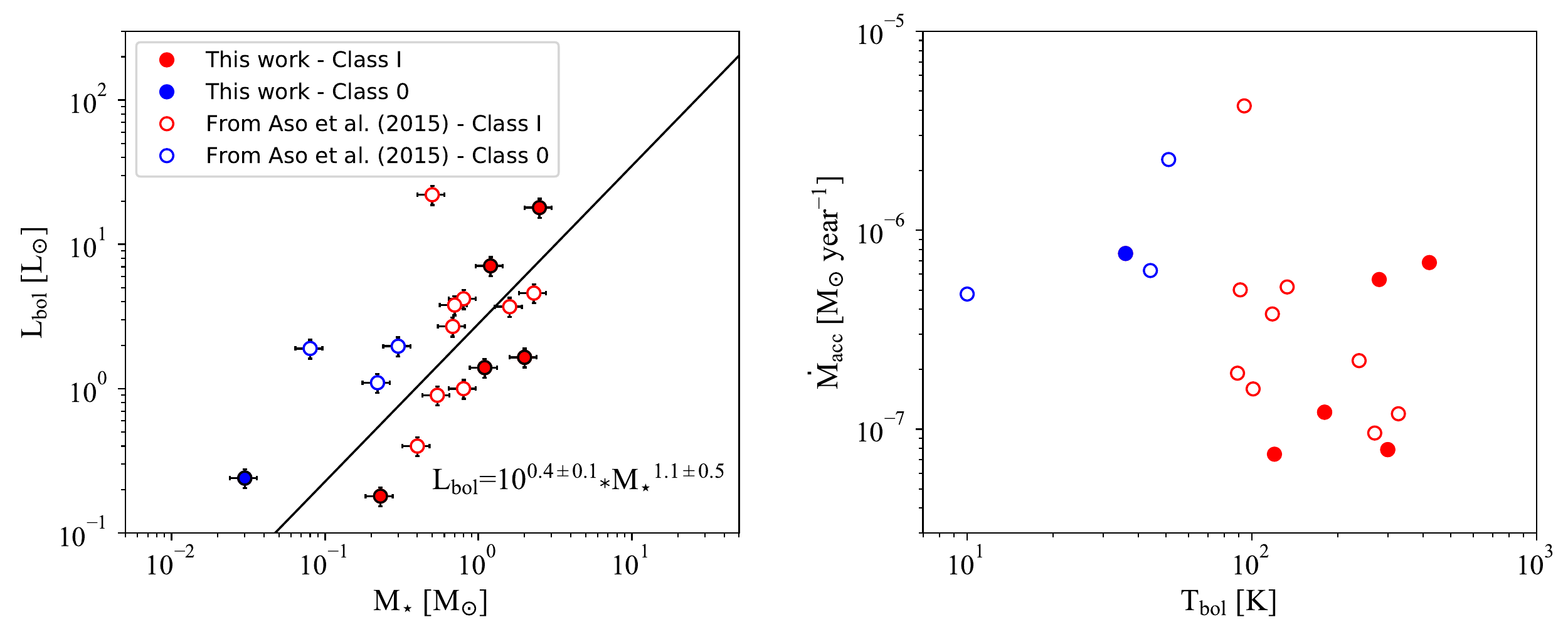}
      \caption[]{\label{fig:L_M}
      \textit{Left:} Bolometric luminosity as a function of stellar mass with data points from this work and from \cite{Aso2015}. The black straight line shows the best fit for all Class~I sources. \textit{Right:} Mass accretion rate as a function of the bolometric temperature.
      }
\end{figure*}

Figure~\ref{fig:M_DES} compares the envelope masses (\textit{M$_\mathrm{env}$}; Table~\ref{table:parameters}), disc masses (\textit{M$_\mathrm{disc}$}; Table~\ref{table:masses}), and stellar masses (\textit{M$_{\star}$}; Table~\ref{table:Fits}). The uncertainties in \textit{M$_\mathrm{env}$} are assumed to be 20$\%$. There is no clear trend between \textit{M$_\mathrm{disc}$} and \textit{M$_{\star}$} or \textit{M$_\mathrm{env}$}, however, all the points satisfy the condition \textit{M$_{\star}$} > \textit{M$_\mathrm{disc}$} and \textit{M$_\mathrm{env}$} > \textit{M$_\mathrm{disc}$}, characteristic of Class~I stages \citep{Robitaille2006}. For \textit{M$_{\star}$} as a function of \textit{M$_\mathrm{env}$},  the stellar mass increases as the mass of the envelope decreases, and only one of the sources (IRAS 16253-2429) satisfies the condition \textit{M$_\mathrm{env}$} > \textit{M$_{\star}$}, which is a characteristic of Class~0 protostars. This is also consistent with its low bolometric temperature, \textit{T$_\mathrm{bol}$} of 36~K, and the infalling profile for the gas kinematics.

The left panel of Fig.~\ref{fig:L_M} shows the bolometric luminosities as a function of stellar masses from Table~\ref{table:Fits} and values of Class~0 and I sources summarised in \cite{Aso2015} (see Table~\ref{table:literature} in the Appendix). The uncertainties in \textit{M$_{\star}$} and \textit{L$_\mathrm{bol}$} are 20$\%$ and 15$\%$, respectively \citep[e.g.][]{Yen2017}, and for the particular case of the binary systems (IRS 43 and IRS 67), the \textit{M$_{\star}$} and \textit{L$_\mathrm{bol}$} values were divided by 2 to account for each component. For Class~I sources, the correlation between \textit{L$_\mathrm{bol}$} and \textit{M$_{\star}$} appears to be linear, with one outlier that corresponds to the source L1551 IRS 5. The straight line in Fig.~\ref{fig:L_M} shows the best fit for all Class~I sources, providing the following power-law relationship:

\begin{equation} 
    L_\mathrm{bol} = 10^{0.4\pm0.1}M_{\star}^{1.1\pm0.5}.
    \label{eq:LM}
\end{equation}

Assuming that \textit{L$_\mathrm{bol}$} results from the gravitational energy that is released by the material accreted onto the surface of the protostar, the mass accretion rate (\textit{$\dot{M}$$_\mathrm{acc}$}) can be estimated from

\begin{equation} 
    \dot{M}_\mathrm{acc} = \frac{L_\mathrm{bol} R_{\star}}{G M_{\star}} \ ,
    \label{eq:Macc}
\end{equation}

\noindent where \textit{R$_{\star}$} is the protostellar radius and \textit{G} the gravitational constant. Assuming \textit{R$_{\star}$}~=~3R$_{\odot}$ \citep{Stahler1980}, the calculated accretion rates are shown in the right panel of Fig.~\ref{fig:L_M} as a function of \textit{T$_\mathrm{bol}$}. Combining Eqs.~\ref{eq:LM} and~\ref{eq:Macc}, a value of (2.4~$\pm$~0.6)~$\times$~10$^{-7}$ M$_{\odot}$~year$^{-1}$ is obtained for \textit{$\dot{M}$$_\mathrm{acc}$} for the Class~I sources, with a minimum and maximum value of 7.5~$\times$~10$^{-8}$ and 7.6~$\times$~10$^{-7}$ M$_{\odot}$~year$^{-1}$, respectively. The mean value is consistent with the mass accretion rates from \cite{Yen2017}, who found \textit{$\dot{M}$$_\mathrm{acc}$} from $\sim$1~$\times$~10$^{-7}$ to 4.4~$\times$~10$^{-6}$ M$_{\odot}$~yr$^{-1}$ for a sample of Class~I sources. The right panel of Fig.~\ref{fig:L_M} shows \textit{$\dot{M}$$_\mathrm{acc}$} as a function of \textit{T$_\mathrm{bol}$}. \cite{Myers1993} defined \textit{T$_\mathrm{bol}$} as the temperature of a blackbody that has the same mean frequency as the observed continuum spectrum, and this is often taken as an indicator of the evolutionary stage of the source \citep[e.g.][]{Dunham2014a, Frimann2016b, Fischer2017}. A tentative decrease in \textit{$\dot{M}$$_\mathrm{acc}$} as the systems evolve is seen in Fig.~\ref{fig:L_M}. Nevertheless, the trend is not clear, and the sample of Class~0 sources is still small.

When the accretion rate is assumed to be constant (2.4~$\times$~10$^{-7}$ M$_{\odot}$~year$^{-1}$), a typical Class~I protostar will reach 1~M$_{\odot}$ in $\sim$4~Myr, which is inconsistent with the derived lifetime for the embedded Class~0/I phase \citep[from 0.13 to 0.78~Myr;][]{Dunham2015, Kristensen2018}. This inconsistency is know as the luminosity problem\textup{} \citep{Kenyon1995}, where a constant accretion rate cannot explain the low luminosities of embedded protostars. With interferometric facilities, the stellar masses are better constrained from molecular line emission, and the only way to reconcile theory with observations is to assume a time-variable accretion rate onto the protostar \citep{Kenyon1995, White2007, Evans2009, Vorobyov2010, Dunham2012, Audard2014, Dunham2014a}. The time-variable or burst model of accretion operates in the embedded phase of protostellar evolution, and the disc spends practically all of its time in a low state (low \textit{$\dot{M}$$_\mathrm{acc}$}) and accretes significant mass in relatively short bursts, which would account for the low average \textit{L$_\mathrm{bol}$} of protostars \citep{Vorobyov2010, Dunham2014a}.

In Fig.~\ref{fig:L_M}, L1551 IRS 5 is the only Class~I source with high \textit{L$_\mathrm{bol}$} and high \textit{$\dot{M}$$_\mathrm{acc}$} that stands out over the others. This source has been proposed to be a young star experiencing an FU Ori-like outburst \citep{Hartmann1996, Osorio2003}. With the exception of L1551 IRS 5, the rest of the Class~I sources in this sample appear to be in a low state of accretion (see Table~\ref{table:literature} in the Appendix).  

If the accretion onto the central star is episodic, the accretion bursts may have strong effects on the chemistry. In consequence, observable chemical effects (such as the C$^{18}$O spatial distribution) can provide clues to the luminosity history \citep[e.g.][]{Jorgensen2013, Jorgensen2015, Frimann2016a}. In our sample, the lack of significantly extended C$^{17}$O emission contrasts with the C$^{18}$O emission observed by \cite{Jorgensen2015} and \cite{Frimann2016a} towards a different sample of Class~0/I sources, which suggests a relatively quiescent phase for our sources or a low C$^{17}$O column density at larger distances from the source.

\section{Chemical evolution}

\subsection{Line emission as a function of \textit{L$_\mathrm{bol}$} and \textit{T$_\mathrm{bol}$}}

The differences in the molecular emission signatures for the sources in the sample may be an indication that the chemistry does depend on the physical evolution of the sources, that is to say, the molecular column densities vary as a function of the source bolometric luminosities and temperatures. For the molecular transitions detected towards the source position, the spectrum was extracted from the pixel that corresponds to the peak of the continuum emission (see Table~\ref{table:coordinates}), and a Gaussian fit was applied to the line profile (see Fig.~\ref{fig:spectra_all} in the Appendix). The resulting values are listed in Table~\ref{table:intensities} and are estimates of the line intensities towards the peak, and not the full integrated emission over the maps. Most of the line profiles show a single central component centred at \textit{V$_\mathrm{source}$}. When more than one component is observed, the intensity listed in Table~\ref{table:intensities} is the sum of all the individual components. For the sources without on-source detection, the 3$\sigma$ upper limit per 1 km~s$^{-1}$, that is, 0.013~Jy~beam$^{-1}$~km~s$^{-1}$, is used in the comparison.

\begin{table}[t]
\caption{Line intensity towards the source position calculated from a Gaussian fit.}
\label{table:intensities}
\centering
\begin{tabular}{l l l l l }
        \hline\hline
                Source                  & \multicolumn{4}{c}{Intensity [Jy beam$^{-1}$ km s$^{-1}$]}\\
                                                & \multicolumn{1}{c}{C$^{17}$O}         & \multicolumn{1}{c}{C$_{2}$H}                    & \multicolumn{1}{c}{H$^{13}$CO$^{+}$}          & \multicolumn{1}{c}{SO$_{2}$}            \\
        \hline
                GSS30-IRS1              & \multicolumn{1}{c}{-}         & \multicolumn{1}{c}{-}   & \multicolumn{1}{c}{-}         & 0.439                                 \\\relax
                [GY92] 30               & 0.081                         & \multicolumn{1}{c}{-}   & \multicolumn{1}{c}{-}         & \multicolumn{1}{c}{-}         \\
                WL 12                   & \multicolumn{1}{c}{-} & \multicolumn{1}{c}{-} & \multicolumn{1}{c}{-}           & \multicolumn{1}{c}{-}         \\\relax
                [GY92] 197              & 0.226                         & 0.073                           & \multicolumn{1}{c}{-}         & \multicolumn{1}{c}{-}         \\
                Elias 29                        & \multicolumn{1}{c}{-} & \multicolumn{1}{c}{-}   & \multicolumn{1}{c}{-}         & 1.036 $^{b}$                          \\
                IRS 43 VLA 1            & 0.049                         & \multicolumn{1}{c}{-}   & 0.133                                 & 0.389 $^{b}$                          \\
                IRS 43 VLA 2            & 0.053                         & \multicolumn{1}{c}{-}   & 0.051                                 & \multicolumn{1}{c}{-}         \\
                IRS 44                  & \multicolumn{1}{c}{-} & \multicolumn{1}{c}{-} & 0.124                                   & 1.478 $^{b}$                          \\
                IRAS 16253-2429         & 0.213                         & 0.206                           & \multicolumn{1}{c}{-}         & \multicolumn{1}{c}{-}         \\
                ISO-Oph 203             & \multicolumn{1}{c}{-} & \multicolumn{1}{c}{-} & \multicolumn{1}{c}{-}           & \multicolumn{1}{c}{-}         \\
                IRS 67 A                        & 0.134                         & \multicolumn{1}{c}{-}   & 0.187                                 & \multicolumn{1}{c}{-}         \\
                IRS 67 B                        & 0.488 $^{a}$                  & 0.077                           & 0.396 $^{a}$                          & 0.221 $^{b}$                            \\
        \hline
\end{tabular}
\tablefoot{$^{(a)}$ Line profile with three components. $^{(b)}$ Line profile with two components.}
\end{table}

\begin{figure}[hbtp]
   \sidecaption
      \includegraphics[width=.24\textwidth]{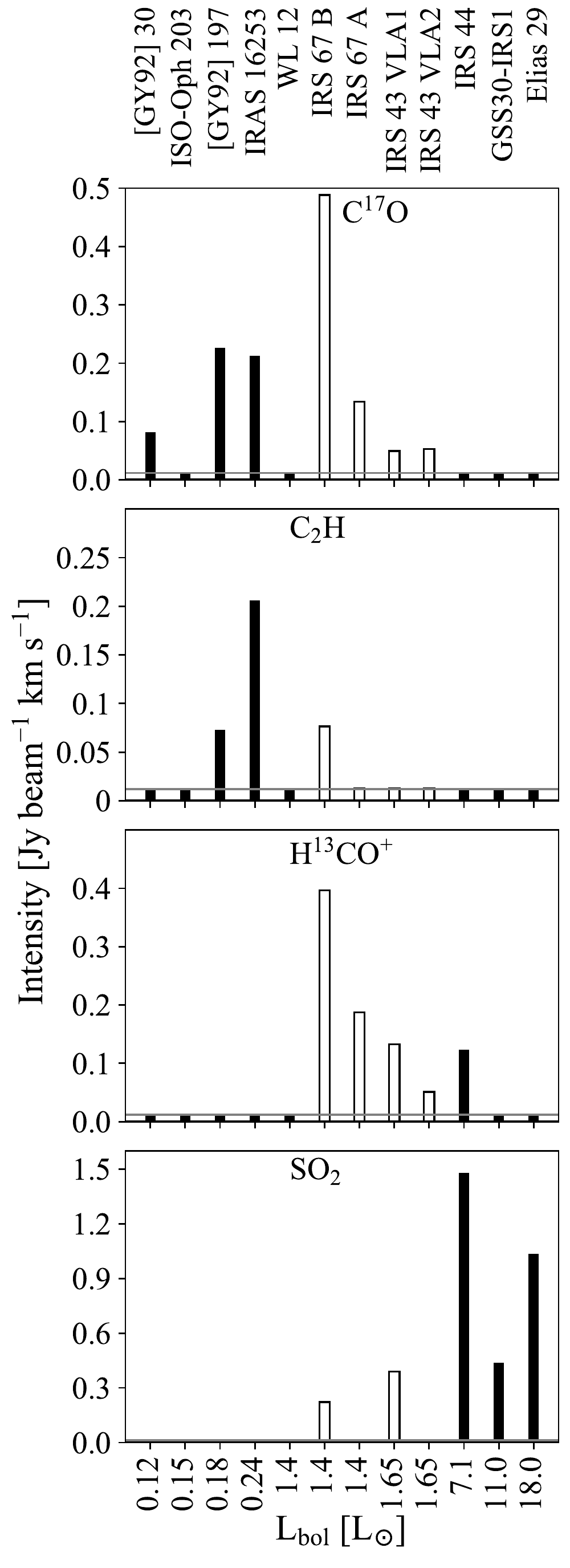}
      \includegraphics[width=.24\textwidth]{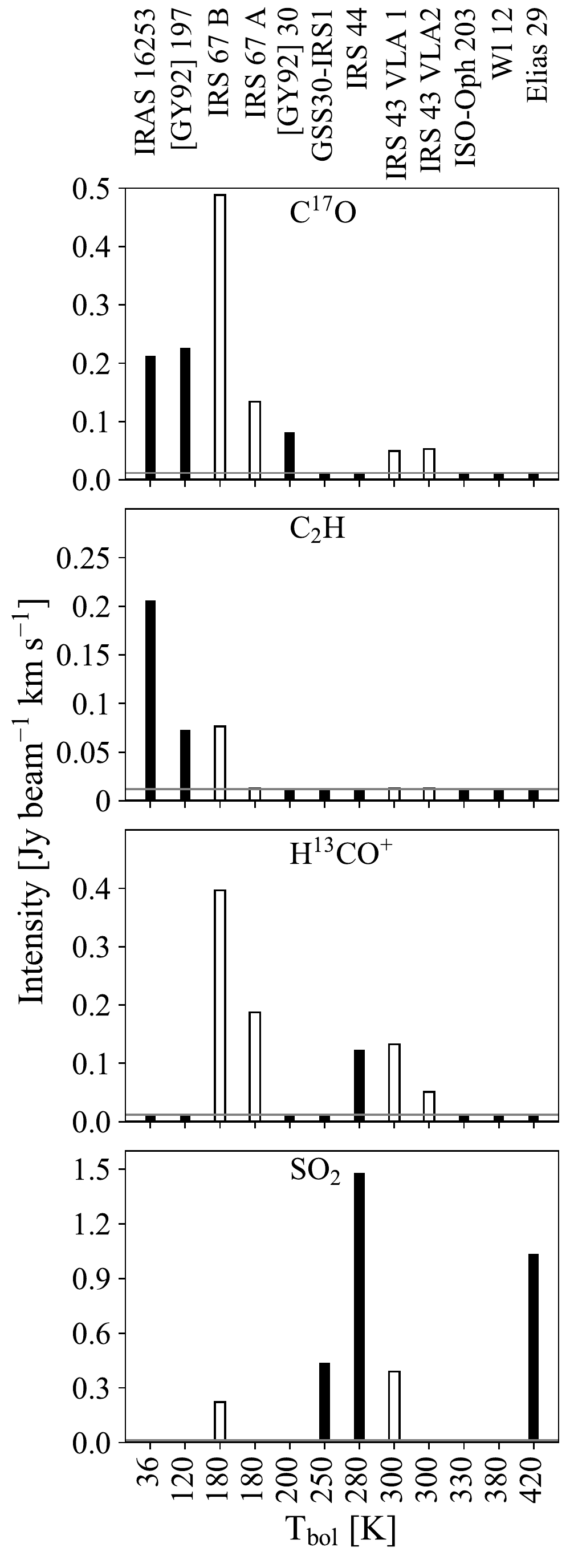}      
      \caption[]{\label{fig:ML}
      Line intensity as a function of the bolometric luminosity \textit{(left)} and bolometric temperature \textit{(right)} towards the source position. The empty bars represent the binary systems, and the grey horizontal line indicates a value of 3$\sigma$.
      }
\end{figure}

Figure~\ref{fig:ML} shows the line intensities from Table~\ref{table:intensities} as a function of \textit{L$_\mathrm{bol}$} and \textit{T$_\mathrm{bol}$}. The binary systems are represented by empty bars in order to distinguish them from the single systems because the former are particularly rich in molecular lines. The binary systems show emission of almost all the transitions presented in Table~\ref{table:observations}, with the exception of CH$_{3}$OH. This chemical richness appears to be related to the mass content and extent of the circumbinary discs, in agreement with the results of \cite{Murillo2018}. 

Considering only the single sources, C$^{17}$O is seen mostly towards the least luminous ones (see upper left panel of Fig.~\ref{fig:ML}), while the opposite situation is the case for SO$_{2}$ (lower left panel of Fig.~\ref{fig:ML}). The sources with high bolometric luminosity are associated with high temperatures close to the protostar. This means that for the same continuum flux, the dust mass will be lower for sources with high \textit{L$_\mathrm{bol}$}, and thus, a lower column density of C$^{17}$O is expected. On the other hand, SO$_{2}$ traces a different physical process than C$^{17}$O and may be related with more energetic processes that are linked to the bolometric luminosity, and thus, the accretion history. If \textit{T$_\mathrm{bol}$} is taken as an evolutionary indicator, C$^{17}$O and C$_{2}$H emission are associated with the less evolved sources. In particular, Figs.~\ref{fig:C17O},~\ref{fig:C2H}, and~\ref{fig:ML} show a trend between the extent of the C$^{17}$O and C$_{2}$H emission and the evolutionary stage of the source. On the other hand, there is no clear correlation between the sources with detected H$^{13}$CO$^{+}$ emission and \textit{L$_\mathrm{bol}$} or \textit{T$_\mathrm{bol}$}.

\begin{figure}[hbtp]
   \centering
      \includegraphics[width=.49\textwidth]{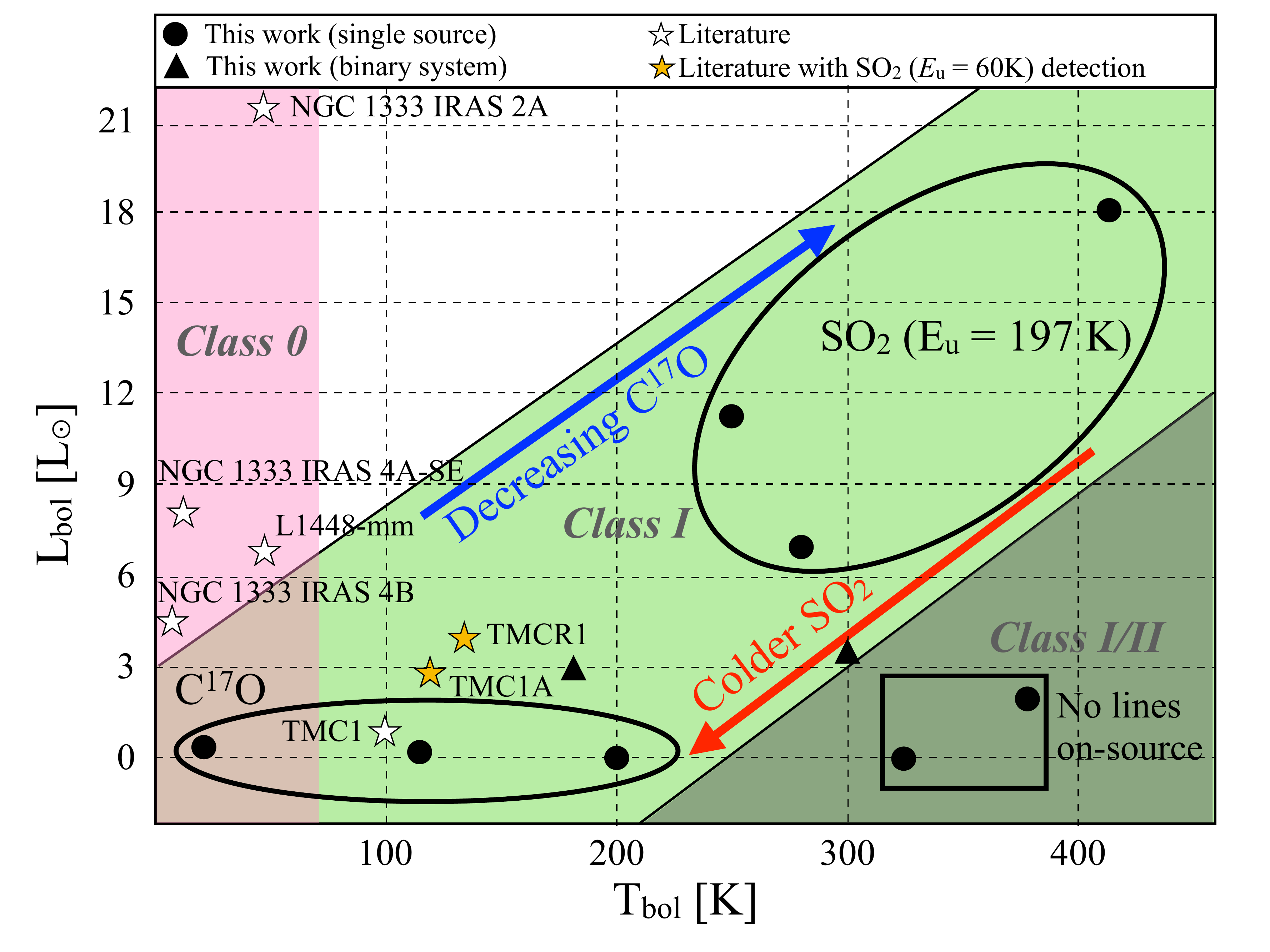}
      \caption[]{\label{fig:LinesTL}
       Bolometric luminosity as a function of the bolometric temperature, highlighting the regions where specific molecular transitions are detected towards the source position, and well-known Class~0 and I sources from \cite{Dunham2015} and  \cite{Harsono2014}, respectively. This plot only covers the lines listed in Table~\ref{table:observations} plus SO$_{2}$ 11$_{1,11}$$-$10$_{0,10}$ from \cite{Harsono2014}. The Class~0 covers \textit{T$_\mathrm{bol}$} $\leq$~70~K \citep{Dunham2014b}, and the Class~I/II region includes the sources where no line emission is detected. 
      }
\end{figure}

Figure~\ref{fig:LinesTL} shows the chemical signatures in a plot of \textit{L$_\mathrm{bol}$} as a function of \textit{T$_\mathrm{bol}$} for the sources that were observed as part of this study with additional points from the literature: Class~0 sources from \cite{Dunham2015} and Class~I sources from \cite{Harsono2014}\footnote{Two Class I sources in the \cite{Harsono2014} study (TMC1A and TMCR1) show SO$_{2}$ 11$_{1,11}$$-$10$_{0,10}$ emission towards the source position, but this transition is from a lower level with \textit{E$_\mathrm{u}$} of 60~K and does not show a broad line profile. Therefore, this cold SO$_{2}$ emission appears to be tracing a different component than the warm SO$_{2}$ seen in Fig.~\ref{fig:SO2}}. Three groups of sources can be identified based on on-source detections. The least evolved and luminous sources show C$^{17}$O emission towards the source position, while the more evolved and luminous sources are associated with SO$_{2}$ emission. In addition, the more evolved sources associated with low luminosities do not show any line detection. As the system evolves from Class~0 to Class~I and later to Class~II, the envelope mass decreases, leading to a lower gas column density and lines may be harder to detect. This is reflected in the emission of a high-density tracer, such as C$^{17}$O: the extent of the emission decreases with the evolutionary stage of the source, followed by a non-detection towards the more evolved sources. For the latter, emission of more abundant isotopologues, such as C$^{18}$O and $^{13}$CO, is expected to be seen, as in Class~II sources. On the other hand, there is a chemical differentiation between C$^{17}$O and SO$_{2}$, where the latter may be tracing a more energetic process that is probably linked to higher accretion rates, which is mostly determined by inner disc properties. Future observations of CO isotopologues and SO$_{2}$ transitions with different \textit{E$_\mathrm{u}$} values towards a larger sample of Class~I sources will provide a more statistical perspective.

\subsection{Absence of warm CH$_{3}$OH emission}

One of the chemical results of this study is the absence of compact CH$_{3}$OH emission towards all of the sources. At large scales, CH$_{3}$OH is found to be in solid form with typical abundances with respect to water, CH$_{3}$OH:H$_{2}$O, of about 5$\%$ \citep{Bottinelli2010, Oberg2011, Boogert2015}. CH$_{3}$OH is expected to sublimate off dust grains in the inner envelopes and in the protostellar discs in regions where the temperature increases above $\sim$90 K \citep{Brown2007}. Such warm CH$_{3}$OH has been inferred for some Class~0 sources based on single-dish observations \citep[e.g.][]{vanDishoeck1995, Schoier2002, Maret2005, Jorgensen2005b, Kristensen2010} and it has also been imaged with interferometers \citep[e.g.][]{Jorgensen2005a, Maury2014, Jorgensen2016, Higuchi2018}.

The absence of CH$_{3}$OH line emission provides strict upper limits to the column densities of the warm gas-phase CH$_{3}$OH. The upper limit to the column density is estimated using the predictions for a synthetic spectrum for methanol calculated under the assumption of LTE and adopting a kinetic temperature of 100~K and a typical line width of 5~km~s$^{-1}$. For the 3$\sigma$ \textit{rms} noise in the spectra of 13 mJy~beam$^{-1}$~km~s$^{-1}$, this corresponds to an upper limit for the column density of 5~$\times$~10$^{14}$~cm$^{-2}$ (assuming that the emission fills the beam uniformly), which is more than four orders of magnitude below that of the Class~0 protostar IRAS~16293-2422 \citep{Jorgensen2016}. As an illustration, Fig.~\ref{fig:CH3OH_spectra} shows the calculated synthetic spectrum for a column density that is higher than this by a factor of five (i.e. 2.5~$\times$~10$^{15}$~cm$^{-2}$).

The low upper limit for the CH$_{3}$OH column density implies that \textit{(i)} there is no hot-core-like region in the inner envelope close to the protostar if its envelope density profile can be extrapolated to the smallest scales and \textit{(ii)} the gas-phase CH$_{3}$OH averaged over the entire disc is low. For the warm gas in the inner envelope, the column density can be translated into a constraint on the abundance by comparing to the results from \cite{Lindberg2014} for the Class~0/I protostar R~CrA~IRS~7B: through line radiative transfer modelling of ALMA detections of the same CH$_3$OH lines, \cite{Lindberg2014} found a CH$_{3}$OH abundance of 10$^{-10}$ in the inner region of the 2.2~M$_{\odot}$ envelope. The average envelope mass for our sample is 0.11~M$_{\odot}$, which is an order of magnitude below that from R CrA IRS 7B, and because our beam size and distance are similar to those of R~CrA~IRS~7B and our upper limits comparable to the brightest lines in the spectra from \cite{Lindberg2014}, the upper limit to the CH$_{3}$OH abundance for our sources should be about an order of magnitude higher than the inferred abundance for R CrA IRS 7B, that is, $\sim$10$^{-9}$. Conversely, by ignoring the envelope, an upper limit to the CH$_{3}$OH abundance for the disc can be estimated by comparing the inferred disc mass from the ALMA dust continuum measurements to the upper limit of the CH$_3$OH column density assuming that both fill the beam uniformly. Adopting an average disc mass of 0.0085~M$_{\odot}$ for our sample and the upper limit to the CH$_{3}$OH column density of 5~$\times$~10$^{14}$~cm$^{-2}$, the corresponding upper limit for the CH$_{3}$OH abundance is 10$^{-10}$, averaged over the entire disc.

These numbers may seem at odds with the results from ice measurements mentioned above, with CH$_{3}$OH at least of order 1$\%$ relative to water, which in turn has a typically quoted abundance ([H$_{2}$O]/[H$_{2}$]) of $10^{-4}$ \citep{Boogert2015}. However, in both cases the assumptions of the physical structures are critical: for example, \cite{Lindberg2014} showed for the protostellar envelope that the CH$_{3}$OH abundance would increase by two orders of magnitude if a constant density profile were used at scales smaller than the disc size (corresponding to the flattening of the inner envelope in rotating collapse), rather than a centrally peaked power-law envelope density profile as expected from free-fall. Because a number of sources in our sample show evidence for resolved discs, it is reasonable to apply the same argument, that is, the upper limit to the CH$_{3}$OH abundance would be 10$^{-7}$ (rather than 10$^{-9}$), which is comparable to the estimates for hot cores towards Class~0 protostars \citep[e.g.][]{Schoier2002,Maret2005,Jorgensen2005b}. Likewise, for the disc argument, \cite{Persson2016} showed that in simple parametrised disc models, only a small fraction, as low as 1$\%$, of the total disc mass may have temperatures above 100~K where water could sublimate. These are the regions where CH$_{3}$OH would also be in the gas phase, thus the upper limit for the CH$_{3}$OH abundance in the warm parts of the discs would likewise be less stringent, $\sim$10$^{-8}$. Furthermore, by analysing the H$_{2}$$^{18}$O emission towards 4 Class~0 sources, \cite{Persson2016} showed that the H$_{2}$O abundance in the warm regions of the discs could be as low as 10$^{-7}$--10$^{-6}$, which would be consistent with the CH$_{3}$OH abundance limit derived above for typical CH$_{3}$OH:H$_{2}$O ice abundance ratios of $\sim$1$\%$.

Obviously, there are still a number of caveats in this analysis, in particular, about the physical structure of the material towards protostars at these small scales. Open questions are for instance the density profiles of the envelopes and the temperature structures of the embedded discs. It can therefore not be ruled out that the absence of CH$_{3}$OH is still to some degree caused by chemical effects such as the suppression of methanol formation that is due to higher temperatures in the precursor environments, as also discussed for the case of Corona Australis by \cite{Lindberg2014}. Future modelling efforts and observation of lower-excited CH$_{3}$OH transitions at large and small scales may provide further insights.

\subsection{Does SO$_{2}$ trace accretion shocks?}

The combination of compact (\textit{r} < 0$\farcs$5 or 70~au; see Fig.\ref{fig:SO2}) high-velocity emission (up to $\pm$10~km~s$^{-1}$; see Fig.\ref{fig:spectra_all}) and the velocity gradient perpendicular to the outflow direction suggests that the warm SO$_{2}$ emission may be tracing warm shocked material, in particular, accretion shocks. Material from the inner envelope falls onto the circumstellar disc and produces accretion shocks at the envelope-disc interface, releasing molecules from the dust grains and altering the chemistry. \cite{Miura2017} investigated the thermal desorption of molecules from the dust-grain surface by accretion shocks and found that the enhancement of some species (such as SO) can be explained by the accretion shock scenario, where the main parameters are the grain size, the pre-shock gas number density, and the shock velocity. Taking a shock velocity of 10~km~s$^{-1}$, \cite{Miura2017} predicted that SO$_{2}$ can be released from the dust-grain surface for a pre-shock gas number density of $\sim$10$^{7}$~cm$^{-3}$. 

Another formation path for SO$_{2}$ is by oxidation of SO in the gas phase \citep{Charnley1997}:

\begin{equation} 
    \mathrm{SO +  OH  \rightarrow SO_{2} + H} ,
    \label{eq:Eq4}
\end{equation}

\noindent which is very efficient for \textit{T} between 100 and 200~K. In this scenario, the SO$_{2}$ abundance depends strongly on the presence of SO and OH in the gas phase. SO can be released from dust grains at lower velocities and densities than SO$_{2}$, since SO has a lower desorption energy (\textit{E$_\mathrm{d0}$}) than SO$_{2}$ \citep[2600~K and 3400~K for SO and SO$_{2}$, respectively;][]{Wakelam2015, Miura2017}. In addition, OH is seen towards Class~I sources and has been associated with shocked regions in the inner envelope close to the protostar \citep{Wampfler2013}. In order to test these two scenarios, oxidation of SO or desorption from dust grains, transitions from warm SO (\textit{E$_\mathrm{u}$} $\sim$200~K) need to be observed. 
 
In addition to SO and SO$_{2}$, CH$_{3}$OH has also been related with shocked regions \citep[e.g.][]{Avery1996, Jorgensen2007}. This raises the question why CH$_{3}$OH is not detected if SO$_{2}$ (or SO) is released from the grain surface by accretion shocks. There are two possibilities: \textit{(i)} CH$_{3}$OH is not released from the grain surface because it has a higher desorption energy (\textit{E$_\mathrm{d0}$} = 5000 K), therefore, a higher pre-shock gas number density is required \citep[$\sim$10$^{8}$~cm$^{-3}$; ][]{Miura2017}, or \textit{(ii)} CH$_{3}$OH is released, but is later destroyed by the high velocities. \cite{Suutarinen2014} demonstrated that CH$_{3}$OH is sputtered from ices in shocks with \textit{v} $\geq$ 3~km~s$^{-1}$, survives at moderate velocities, but is later dissociated for \textit{v} $\geq$ 10~km~s$^{-1}$.

A Keplerian disc or disc winds are less plausible scenarios to explain the SO$_{2}$ emission. For a Keplerian profile, high velocities ($\geq$ 5~km~s$^{-1}$) would be reached at scales smaller than 0$\farcs$2 (see Fig.\ref{fig:PV1}), therefore, the high-velocity SO$_{2}$ component seen at $\sim$0$\farcs$5 (see Fig.\ref{fig:SO2}) is not consistent with a Keplerian profile. For disc winds towards Class~I sources, it has been shown that species such as SO do not survive beyond $\sim$1~au \citep{Panoglou2012}. In contrast, towards Class~0 sources, a disc wind driven by SO survives between 10 and 100~au \citep{Panoglou2012}, in agreement with what was found towards the Class~0 source HH212: \cite{Tabone2017} proposed that SO and SO$_{2}$ are tracing a disc wind and the emission is observed between $\sim$50 and $\sim$150~au. In any case, neither option can be completely ruled out, and higher spatial resolution is required in order to create a PV diagram and obtain a velocity profile for SO$_{2}$.

\section{Summary}

We presented high angular resolution (0$\farcs$4, $\sim$60~au) ALMA observations of 12 Class~I sources in the Ophiuchus star-forming region. The continuum emission at 0.87 mm was analysed together with C$^{17}$O, C$^{34}$S, H$^{13}$CO$^{+}$, SO$_{2}$, C$_{2}$H, and CH$_{3}$OH, and the main results are provided below.

Of the 12 sources, 2 show no continuum emission nor molecular line emission, whereas another 2 show continuum emission but no line detection. Two sources are proto-binary systems with very rich line emission, and the remaining 6 sources show continuum emission plus some of the molecular transitions. C$^{17}$O is seen towards the less evolved sources and the binary systems, tracing high gas column densities. Keplerian profiles are found for three sources, while an infalling profile is seen for one of them. More abundant isotopologues, such as C$^{18}$O and $^{13}$CO, may be better disc tracers for the more evolved sources.

The non-detection of warm CH$_{3}$OH implies that there in no hot-core-like region in the inner envelope close to the protostar (which follows a power-law density profile) and that the averaged CH$_{3}$OH column density over the entire disc is low. This suggests that \textit{(i)} the presence of a disc flattens the envelope density profile at small scales and thus leads to a low column density of warm material, \textit{(ii)} only a small portion (1$\%$) of the disc may have temperatures above 100~K, or \textit{(iii)} chemical effects may suppress the formation of methanol. Clearly, future modelling efforts and observations of colder CH$_{3}$OH at envelope and disc scales are required in order to provide stronger conclusions.  

Warm (\textit{E$_\mathrm{u}$} = 197~K) and compact (\textit{r} < 70~au) SO$_{2}$ emission  is detected towards five of the sources, with particularly large line widths (between $-$10 and 10~km~s$^{-1}$) towards Elias 29 and IRS 44. This emission may be related with accretion shocks. The shocks would also liberate CH$_{3}$OH from dust mantles, but it would later be destroyed by the high velocities (> 10~km~s$^{-1}$). 

The fact that C$^{17}$O is detected towards the less evolved and less luminous sources agrees with a decrease in gas column density, which is a consequence of the evolution of the system, and with the low column density of material due to high temperatures. The envelope mass decreases as the system evolves, therefore the gas column density related to quiescent material decreases and lines are hardly detected. However, no similar trend is observed for SO$_{2}$ , and instead, a chemical differentiation between C$^{17}$O and SO$_{2}$ is seen. SO$_{2}$ is detected towards the most evolved sources with high \textit{L$_\mathrm{bol}$}, and is therefore related with higher accretion rates and a different physical process. 

The comparison between disc, envelope, and stellar masses shows a trend between \textit{M$_{\star}$} and \textit{M$_\mathrm{env}$}: the most massive stars are related with less envelope material, as expected. In addition, \textit{L$_\mathrm{bol}$} shows a linear dependence with \textit{M$_{\star}$} for Class~I sources, where the best fit gives \textit{L$_\mathrm{bol}$} = 10$^{0.4\pm0.1}$\textit{M$_{\star}$}$^{1.1\pm0.5}$. Assuming that \textit{L$_\mathrm{bol}$} is a consequence of accretion onto the protostar, a mean \textit{$\dot{M}$$_\mathrm{acc}$} of (2.4~$\pm$~0.6)~$\times$~10$^{-7}$ M$_{\odot}$~year$^{-1}$ is calculated, with values ranging from 7.5~$\times$~10$^{-8}$ to 7.6~$\times$~10$^{-7}$ M$_{\odot}$~year$^{-1}$. If \textit{$\dot{M}$$_\mathrm{acc}$} is constant, the time required to accrete enough mass will be greater that the mean lifetime of the embedded stage, supporting the scenario of episodic accretion bursts and a variable accretion rate. Within this scenario, the Class~I sources discussed in this work may be in a quiescent phase, with the exception of L1551 IRS 5.

This work shows the importance of a representative sample for exploring the physical and chemical structure of Class~I sources, by comparing not only the continuum emission, but also the emission of specific molecules and the protostellar masses obtained from the velocity profiles. The observation of disc and warm gas tracers is crucial in order to interpret the physical and chemical processes and evolution at disc scales. Future observations will provide more statistical results, and the study of other species will contribute to a better understanding of the chemical evolution of low-mass protostars.

\begin{acknowledgements}

We thank the anonymous referee for a number of good suggestions that helped us to improve this work. This paper makes use of the following ALMA data: ADS/JAO.ALMA\#2013.1.00955.S. ALMA is a partnership of ESO (representing its member states), NSF (USA) and NINS (Japan), together with NRC (Canada), NSC and ASIAA (Taiwan), and KASI (Republic of Korea), in cooperation with the Republic of Chile. The Joint ALMA Observatory is operated by ESO, AUI/NRAO and NAOJ. The group of JKJ acknowledges support from the European Research Council (ERC) under the European Union's Horizon 2020 research and innovation programme (grant agreement No 646908) through ERC Consolidator Grant ``S4F''. Research at the Centre for Star and Planet Formation is funded by the Danish National Research Foundation. 

\end{acknowledgements}

\bibliographystyle{aa} 
\bibliography{References}

\begin{appendix}

\section{CH$_{3}$OH emission towards [GY92] 30}

\begin{figure*}[ht]
   \centering
      \includegraphics[width=.9\textwidth]{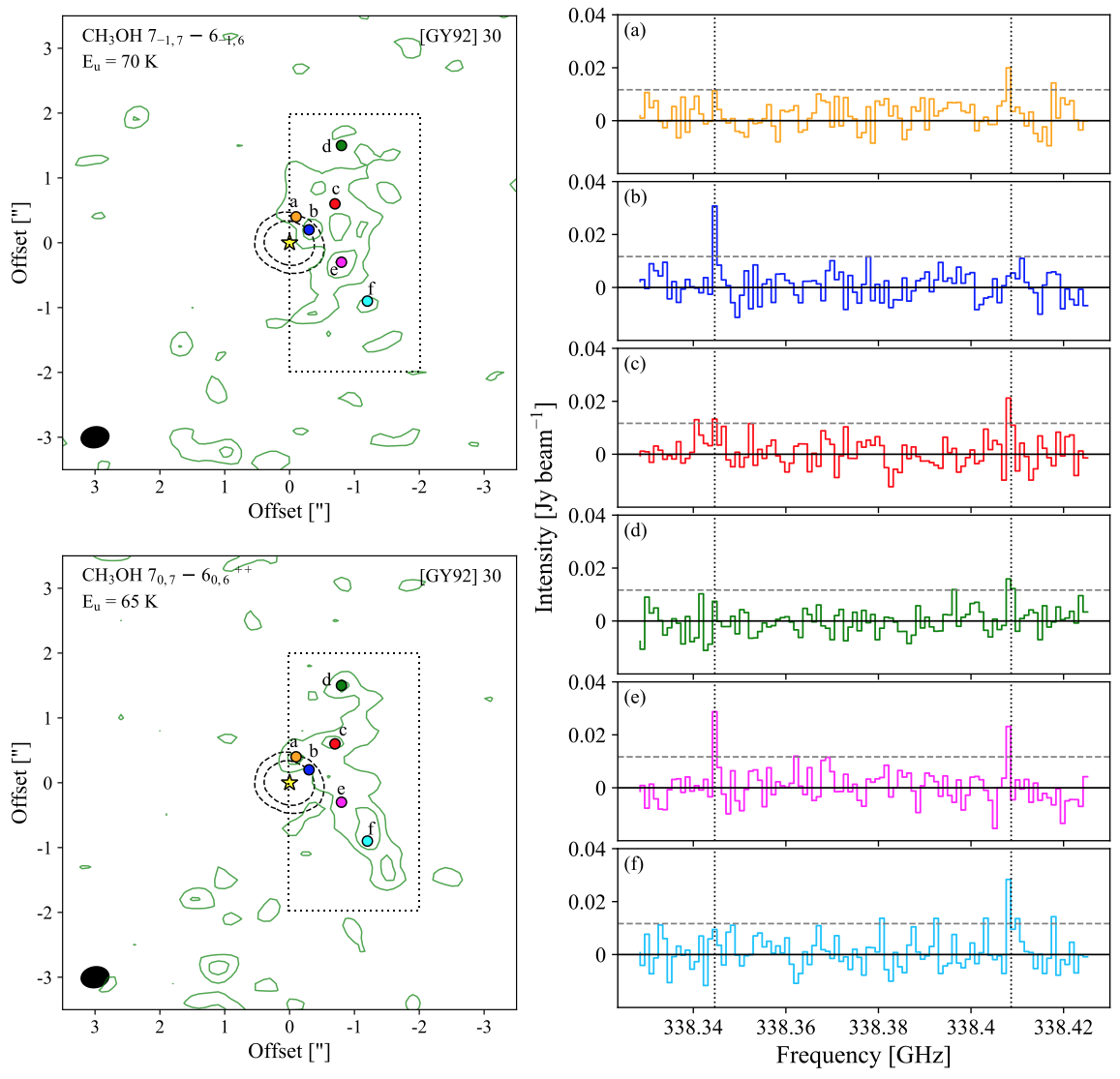}
      \caption[]{\label{fig:CH3OH}
      \textit{Left}: Contour maps of CH$_{3}$OH 7$_{-1}$$-$6$_{-1}$ E and CH$_{3}$OH 7$_{0}$$-$6$_{0}$ A$^{+}$ towards [GY92] 30. The contours start at 3$\sigma$ and follow a step of 3$\sigma$ ($\sigma$~=~4~mJy~beam$^{-1}$~km~s$^{-1}$), representing velocities from -0.5 to 0.5 km~s$^{-1}$. The dashed black contours show the continuum emission for values of 4 and 25$\sigma$. The yellow star indicates the position of the 2D Gaussian fit, and the synthesised beam is represented by the black filled ellipse. The dotted box represents the region from which the spectrum of Fig.~\ref{fig:CH3OH_bis} is integrated. \textit{Right}: Spectra towards different positions marked on the contour maps. The grey dashed horizontal line represents a value of 3$\sigma,$ and all the spectra are rebinned by a factor of 4. The dotted black vertical lines indicate the rest frequency of the two CH$_{3}$OH transitions.
      }
\end{figure*}

\begin{figure}[!htb]
   \centering
      \includegraphics[width=.49\textwidth]{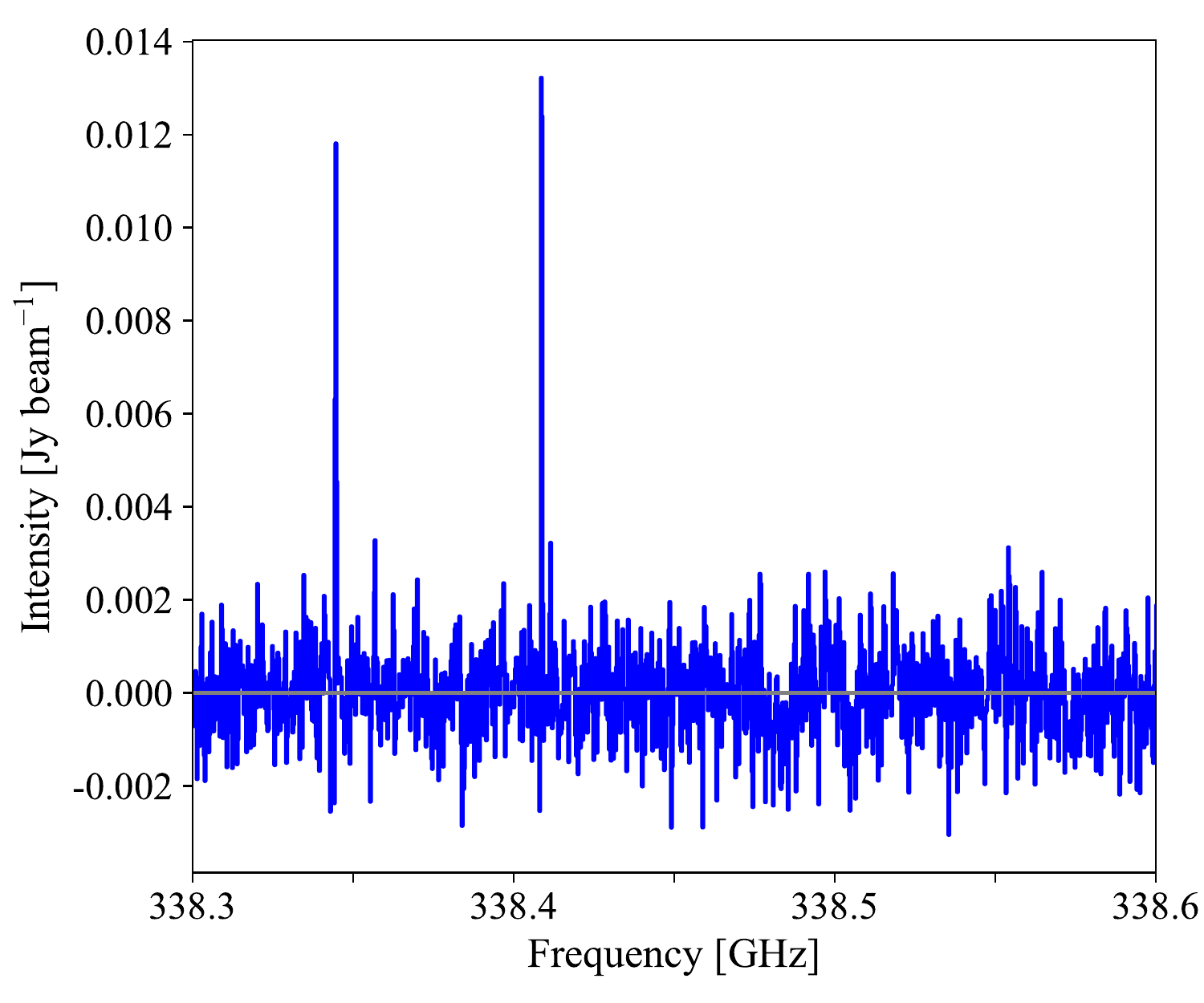}
      \caption[]{\label{fig:CH3OH_bis}
      Integrated spectrum towards the squared region highlighted in the left panels of Fig.~\ref{fig:CH3OH}.
      }
\end{figure}

Two CH$_{3}$OH transitions are detected towards [GY92] 30, where the emission peaks beyond the 25$\sigma$ continuum contour and no emission is seen towards the source position (Fig.~\ref{fig:CH3OH}). These transitions are associated with the lowest \textit{E$_\mathrm{u}$} (65 and 70~K) and the emission is related with low velocities of between $-$0.5 and 0.5~km~s$^{-1}$ from the source velocity. In addition, they present different morphologies, and the spectra taken towards different positions show a variation in the intensity of both lines. The integrated spectrum towards a region of $\sim$300$\times$600~au is shown in Fig.~\ref{fig:CH3OH_bis}, where both CH$_{3}$OH transitions are detected. This emission may be related with quiescent envelope material because [GY92] 30 has the more massive envelope from the sample (see Table~\ref{table:parameters}).

\section{Mass accretion rates}

The values of \textit{L$_\mathrm{bol}$}, \textit{T$_\mathrm{bol}$}, \textit{M$_{\star}$}, and \textit{$\dot{M}$$_\mathrm{acc}$}, plotted in Fig.~\ref{fig:L_M}, are listed in Table~\ref{table:literature}. \textit{L$_\mathrm{bol}$}, \textit{T$_\mathrm{bol}$}, and \textit{M$_{\star}$} are taken from this work and from the literature (see Tables~\ref{table:parameters} and \ref{table:Fits}), while \textit{$\dot{M}$$_\mathrm{acc}$} is calculated from Eq.~\ref{eq:Macc}.

\begin{table*}[hb]
\caption{Bolometric luminosity, bolometric temperature, and stellar masses of Class~0 and I sources from this work and from the literature, with the calculated mass accretion rates.}
\label{table:literature}
\centering
\begin{tabular}{l c c c c l}
        \hline\hline
                Source          & \enspace\textit{L$_\mathrm{bol}$}     & \textit{T$_\mathrm{bol}$}       & \enspace\textit{M$_{\star}$}          & \textit{$\dot{M}$$_\mathrm{acc}$}                       & References            \\
                                        & \enspace[L$_{\odot}$]                         & [K]                                             & \enspace[M$_{\odot}$]                 & [10$^{-6}$~M$_{\odot}$~yr$^{-1}$]               &                               \\
        \hline
                                                                                        \multicolumn{6}{c}{Class~0}                                                                                                             \\
        \hline
                NGC 1333 IRAS 4A2       & \enspace1.9           & \enspace51            & \enspace0.08            & 2.27  & \cite{Choi2010}                                               \\
                VLA 1623A                       & \enspace1.1           & \enspace10              & \enspace0.22          & 0.48  & \cite{Murillo2013}                                            \\
                L1527 IRS                       & \quad1.97             & \enspace44              & \enspace0.30          & 0.63  & \cite{Ohashi2014}                                             \\
                IRAS 16253-2429         & \quad0.24             & \enspace36            & \enspace0.03            & 0.76  & This work                                                     \\
        \hline
                                                                                        \multicolumn{6}{c}{Class~I}                                                                                                             \\
        \hline
                R CrA IRS 7B                    & \enspace4.6           & \enspace89              & 2.3                           & 0.19  & \cite{Lindberg2014}                                   \\
                L1551 NE                                & \enspace4.2           & \enspace91              & 0.8                           & 0.50  & \cite{Froebrich2005,Takakuwa2014b}            \\
                L1551 IRS 5                     & 22.1                  & \enspace94              & 0.5                           & 4.22  & \cite{Kristensen2012,Chou2014}                        \\
                TMC1                            & \enspace0.9           & 101                     & \enspace0.54          & 0.16  & \cite{Harsono2014}                                    \\
                TMC-1A                          & \enspace2.7           & 118                     & \enspace0.68          & 0.38  & \cite{Aso2015}                                                \\
                TMR1                            & \enspace3.8           & 133                     & 0.7                           & 0.52  & \cite{Harsono2014}                                    \\
                L1489 IRS                       & \enspace3.7           & 238                     & 1.6                           & 0.22  & \cite{Yen2014}                                                \\
                L1536                           & \enspace0.4           & 270                     & 0.4                           & 0.09  & \cite{Harsono2014}                                    \\
                IRS 63                          & \enspace1.0           & 327                     & 0.8                           & 0.12  & \cite{Kristensen2012,Brinch2013}                      \\\relax
                [GY92] 197                      & \quad0.18             & 120                     & \enspace0.23          & 0.07  & This work                                                     \\
                Elias 29                                & 18.0                  & 420                     & 2.5                           & 0.69  & This work                                                     \\
                IRS 43 VLA 1                    & \quad1.65             & 300                     & 0.9                           & 0.17  & This work                                                     \\                                                              IRS 43 VLA 2                        & \quad1.65             & 300                   & 0.9                             & 0.17  & This work                                                     \\
                IRS 44                          & \enspace7.1           & 280                     & 1.2                           & 0.56  & This work                                                     \\
                IRS 67 A                                & \enspace1.4           & 180                     & 1.1                           & 0.12  & This work                                                     \\
                IRS 67 B                                & \enspace1.4           & 180                     & 1.1                           & 0.12  & This work                                                     \\
        \hline
\end{tabular}
\end{table*}

\section{Gaussian fits}

\begin{figure*}[h]
   \centering
      \includegraphics[width=.97\textwidth]{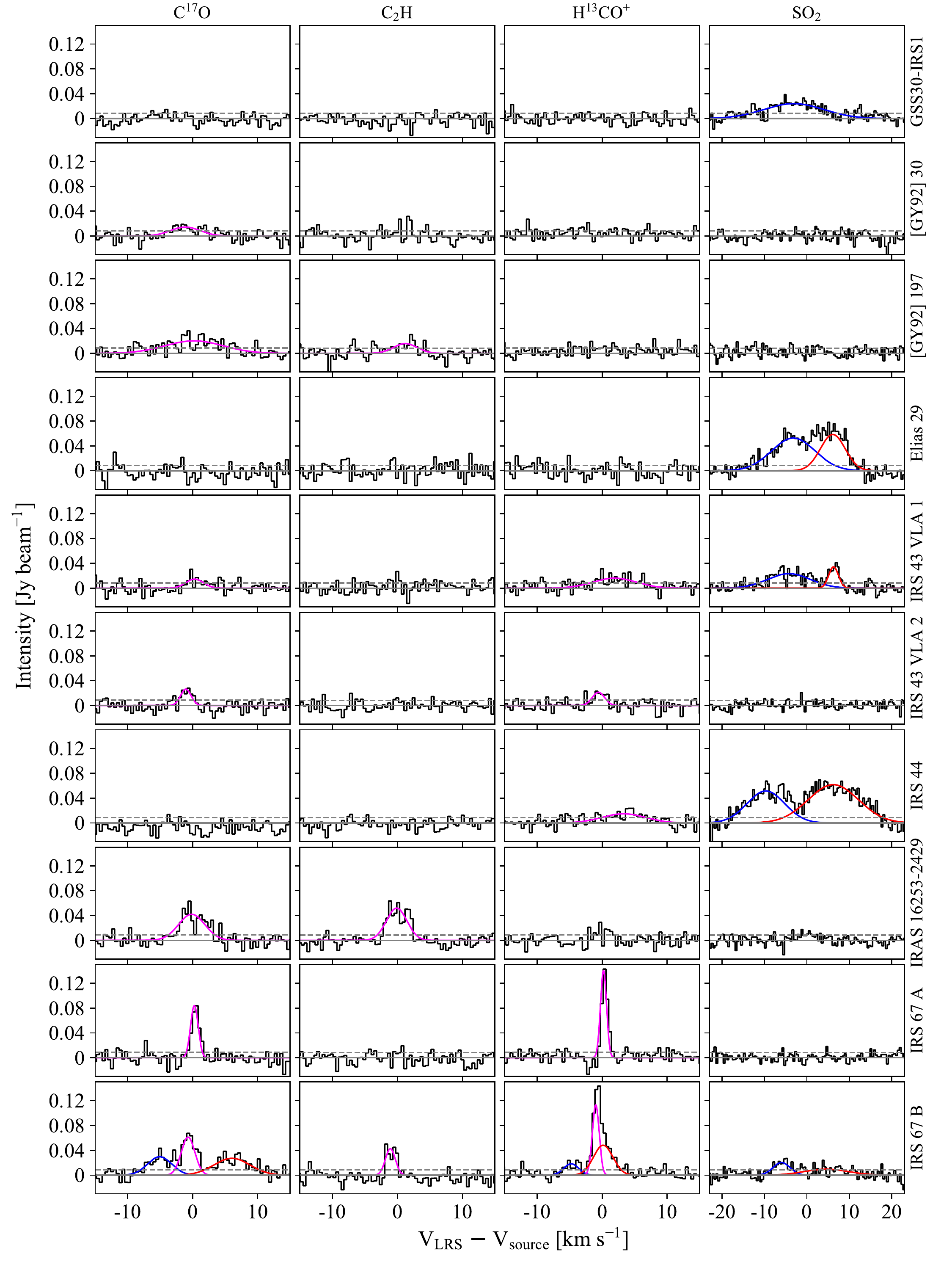}
      \caption[]{\label{fig:spectra_all}
      Spectra towards the source position; a Gaussian fit is overplotted (see Table~\ref{table:intensities}). The blue, magenta, and red curves represent blue-shifted, quiescent, and red-shifted material, respectively. The dashed grey horizontal line shows the value of 3$\sigma$. All spectra have been shifted to the source velocity (see Table~\ref{table:molecules}).
      }
\end{figure*}

The spectra extracted from the source position (see Table~\ref{table:coordinates}) were fitted by a Gaussian profile with one, two, or three components (see Fig.~\ref{fig:spectra_all}). The resulting intensities from the fit are listed in Table~\ref{table:intensities} and plotted in Fig.~\ref{fig:ML}. Most of the line profiles show a single central component centred at \textit{V$_\mathrm{source}$}, with the exception of source IRS 67 B and the SO$_{2}$ transition, which show more than one component. C$^{17}$O and H$^{13}$CO$^{+}$ emission towards IRS 67 B shows three components that are associated with blue-shifted, red-shifted, and a central component associated with more quiescent material, while the SO$_{2}$ emission shows two components related with blue-shifted and red-shifted material (with the exception of GSS30-IRS1, where only blue-shifted emission is seen).

\end{appendix}

\end{document}